# Numerical modelling of flame spread over thin circular ducts

Vipin Kumar[1], Kambam Naresh[1], Amit Kumar[1]*

[1,1]*Department of Aerospace Engineering, Indian Institute of Technology Madras, India

**Abstract**

This paper presents a numerical investigation into the phenomenon of flame spread over thin circular ducts in normal gravity and microgravity environments. Flame spread over such geometry is of significant interest due to its relevance in various practical applications, including tubes for flow purpose in medical system, fire safety in spacecrafts, ducts as well as wiring tubes. This study comprises of a comprehensive investigation of key parameters affecting flame spread rate, including fuel radius and opposed flow speed in normal gravity and microgravity environments. A 2-D axisymmetric flame spread model accounted for char and numerical simulations were performed which revealed valuable insights into the underlying mechanisms governing flame spread over such geometry. The results computed from the numerical model is compared with the experimentally observed flame spread rate to validate the numerical model which can be used to gain a comprehensive understanding of the underlying physical phenomena.

As the radius of circular duct increases the flame spread rate increases both in normal gravity and microgravity environments. The conduction heat feedback and radiation heat gain coming from hot char through gas phase at inner core region are the two major mechanisms which controls the flame spread phenomena over the circular duct fuels. The flame spread rate at different flow ranging from quiescent (0 cm/s) to 30 cm/s is also evaluated and 21 % oxygen and found a non-monotonic increasing decreasing trend of flame spread rate at different opposed flow speed in both normal gravity and microgravity environments.

# Nomenclature

| | | |
|---|---|---|
| $\bar{A}_s$ | : | Virgin fuel pre-exponential factor ($= 4.9 \times 10^{\,9}$ /s) |
| $A_s$ | : | Non-dimensional virgin fuel pre-exponential factor ($=\bar{A}_s/\bar{U}_R$) |
| $A_{CS}$ | : | Non-dimensional cross-sectional area of the hollow fuel ($\pi(R_o^2 - R_i^2)$) |
| $A_{surf,i}$ | : | Non-dimensional inner surface area per unit length of the fuel $(2\pi R_i)$ |
| $A_{surf,o}$ | : | Non-dimensional outer surface area per unit length of the fuel $(2\pi R_o)$ |
| $B_O$ | : | Boltzmann number $(= \rho^* c_p^* \bar{U}_R / (\sigma \bar{T}_\infty^3))$ |
| $\bar{B}_g$ | : | Gas-phase pre-exponential factor $(=1.4 \times 10^{12}\ mm^3/(kg\text{-}s))$ |
| $c_p$ | : | Non-dimensional gas- phase specific heat $(= \bar{c}_p / c_p^*)$ |
| $\bar{c}_p$ | : | Gas- phase specific heat $= \sum_{i=1}^{N} \bar{c}_{p,i} Y_i$ |
| $c_p^*$ | : | Reference gas- phase specific heat $(=1.38 \times 10^3\ kJ/kgK)$ |
| $\bar{c}_s$ | : | Solid- phase specific heat |
| $c_s$ | : | Non-dimensional specific heat of solid $(= \bar{c}_s / c_p^*)$ |
| $Da$ | : | Damköhler number $(= \alpha^* \rho^* \bar{B}_g / \bar{U}_R^2)$ |
| $D_i$ | : | Non-dimensional diffusion coefficient of species $i$ |
| | | $(\rho D_i / \rho^* D_i^*) = (T/T^*)^{0.7}$ *(Sutherland law)* |
| $\bar{E}_g$ | : | Gas- phase activation energy $(=9.5 \times 10^4\ kJ/kg.mo.)$ |
| $E_g$ | : | Non-dimensional gas- phase activation energy $(= \bar{E}_g / (\bar{R}_u \bar{T}_\infty))$ |
| $\bar{E}_s$ | : | Solid -phase activation energy $(=5.8 \times 10^5\ kJ/kgmol)$ |
| $E_s$ | : | Non-dimensional solid phase activation energy $(= \bar{E}_s / (\bar{R}_u \bar{T}_\infty))$ |
| $f_i$ | : | Stoichiometric mass ratio of species $i$ / fuel in gas phase |
| $F_i$ | : | Stoichiometric ratio of species $i$/char in char oxidation reaction |
| $\bar{g}$ | : | Gravitational acceleration |
| $\bar{g}_e$ | : | Gravitational acceleration on the surface of earth $(\bar{g}_e = 9.81\ m/s^2)$ |
| $g$ | : | Non-dimensional gravitational acceleration $(\bar{g}/\bar{g}_e)$ |
| $G$ | : | Total incident radiation |
| $h_i$ | : | Non-dimensional enthalpy of species $i$ $[(\bar{h}_i^o + \int_{\bar{T}_o=298\ K}^{\bar{T}} \bar{c}_{p,i} d\bar{T}) / c_p^* \bar{T}_\infty]$ |
| $\Delta \bar{H}_R^o$ | : | Heat of combustion (= $16.7 \times 10^3$ kJ/kg) |
| $\Delta H_R^o$ | : | Non-dimensional heat of combustion (= $\Delta \bar{H}_R^o / (c_p^* \bar{T}_\infty)$ = 40.4) |
| $I$ | : | Intensity of gas radiation |
| $I_b$ | : | Blackbody intensity at local temperature |
| $k$ | : | Non-dimensional gas thermal conductivity (= $\bar{k} / k^*$) |
| $k^*$ | : | Reference gas thermal conductivity (= $4.6 \times 10^{-6}$ kJ/(m-s-K)) |
| $K$ | : | Absorption coefficient of the medium |
| $L$ | : | Non-dimensional latent heat of pyrolysis (= 0) |
| $Le_i$ | : | Lewis number of species $i$ ($Le_F = 1, Le_{O2} = 1.11, Le_{CO2} = 1.39, Le_{H2O} = 0.83,$) |
| $\bar{L}_R$ | : | Reference length (gas-phase thermal length, ($\alpha^*/\bar{U}_R$) |
| $\bar{\dot{m}}''_s$ | : | Mass flux from solid (= $\bar{A}_s \bar{\rho}_s \exp(-E_s/T_s)$) |
| $\dot{m}''_s$ | : | Non-dimensional mass flux from solid (= $\bar{\dot{m}}''_s / (\rho^* \bar{U}_R)$) |
| $\dot{m}'''_{char}$ | : | Non-dimensional mass of char per unit volume (= $\bar{\dot{m}}'''_{char} / (\rho^* \bar{U}_R / \bar{L}_R)$) |
| $p$ | : | Non-dimensional pressure (= $(\bar{p} - \bar{p}_\infty)/(\rho^* \bar{U}_R^2)$) |
| $\bar{p}_\infty$ | : | Ambient pressure (= 1 atm.) |
| $\vec{\dot{q}}_r$ | : | Gas radiation heat flux |
| $q_{c,i}''$ | : | Non-dimensional conduction heat flux from gas to inner surface of the fuel (= $\overline{q_{c,i}''}/c_p^* \rho^* \bar{U}_R$) |
| $q_{c,o}''$ | : | Non-dimensional conduction heat flux from gas to outer surface of the fuel (= $\overline{q_{c,o}''}/c_p^* \rho^* \bar{U}_R$) |
| $q_{r,i}''$ | : | Net radiation heat flux at inner surface of the fuel (= $\bar{q}''_{r,i}/\sigma \bar{T}_\infty^4$) |
| $q_{r,o}''$ | : | Net radiation heat flux at outer surface of the fuel (= $\bar{q}''_{r,o}/\sigma \bar{T}_\infty^4$) |
| $q_r^{r+}, q_r^{r-}$ | : | Positive and negative component of $q_r$ in r-direction |
| $q_r^{x+}, q_r^{x-}$ | : | Positive and negative component of $q_r$ in x-direction |
| $r$ | : | Non-dimensional r-coordinate $(\bar{r}/\bar{L}_R)$ |
| $R$ | : | Circular duct radius |
| $r_f$ | : | Non-dimensional radius of the fuel $(= R/\bar{L}_R)$ |
| $\bar{R}_u$ | : | Universal gas constant $(=8.305\ kJ/(kg.mol\text{-}K))$ |
| $Re$ | : | Reynolds number $(=\rho^*\bar{U}_R\bar{L}_R/\mu^*)$ |
| $T^*$ | : | Reference temperature $(125°K)$ |
| $T$ | : | Non-dimensional gas-phase temperature $(\bar{T}/\bar{T}_\infty)$ |
| $T_L$ | : | Non-dimensional temperature at which latent heat is given $(=\bar{T}_L/\bar{T}_\infty)$ |
| $T_s$ | : | Non-dimensional solid-phase temperature $(=\bar{T}_s/\bar{T}_\infty)$ |
| $\bar{T}_\infty$ | : | Ambient temperature $(= 300K)$ |
| $t$ | : | Non dimensional time $(=\bar{t}/t_R)$ |
| $t_R$ | : | Reference time $(=\alpha^*/\bar{U}_R^2)$ |
| $\bar{u}$ | : | Velocity in x-direction |
| $u$ | : | Non-dimensional velocity in x-direction $(=\bar{u}/\bar{U}_R)$ |
| $\bar{U}_B$ | : | Reference buoyant velocity $[g_R\beta_R(T_\infty - T_F)\alpha^*]^{1/3}$ |
| $\bar{U}_R$ | : | Reference velocity $(= max\ (\bar{U}_\infty + \bar{U}_B),\ 50\ mm/s)$ |
| $\bar{U}_\infty$ | : | Forced free stream flow velocity |
| $\bar{v}$ | : | Velocity in r-direction |
| $v$ | : | Non-dimensional velocity in r-direction $(=\bar{v}/\bar{U}_R)$ |
| $v_w$ | : | Non-dimensional blowing velocity in r-direction at the solid surface $(=\bar{v}_w/\bar{U}_R)$ |
| $V$ | : | Non-dimensional velocity vector $(= u\ \hat{\imath} + v\ \hat{\jmath})$ |

| $\forall$ | : | Non-dimensional volume of a cell $(=\overline{\forall}/L_R^3)$ |
|---|---|---|
| $V_f$ | : | Flame spread rate |
| $v_f$ | : | Non-dimensional flame spread rate $(= V_f/\overline{U}_R)$ |
| $x$ | : | Non-dimensional x-coordinate $(\bar{x}/\overline{L}_r)$ |
| $Y_i$ | : | Mass fraction of species $i$ $(i = F, O_2, CO_2, H_2O)$ |
| $Y_{pyro}$ | : | Mass fraction of pyrolysate $(=0.85)$ |
| $Y_{char}$ | : | Mass fraction of char |

**Greek Symbols**

| $\alpha^*$ | : | Reference gas thermal diffusivity $(2.13 \times 10^{-4}\ m^2/s)$ |
|---|---|---|
| $\beta$ | : | Extinction coefficient $(= k + \sigma_s)$ |
| $\sigma_s$ | : | Scattering coefficient |
| $\Phi$ | : | Scattering phase function |
| $\mu$ | : | Non-dimensional gas dynamic viscosity $(= \bar{\mu}/\mu^*)$ |
| $\mu^*$ | : | Reference gas viscosity $(4.1 \times 10^{-5}\ kg/m/s)$ |
| $\dot{\omega}_i'''$ | : | Sink/source term in gas-phase reaction |
| | : | $(= f_i \dot{\omega}_F''' = f_i Da \rho^2 Y_F Y_{O2} \exp(-E_g/T))$ |
| $\rho$ | : | Non-dimensional gas density |
| $\rho^*$ | : | Reference gas density $(0.275\ kg/m^3)$ |
| $\rho_{char}$ | : | Non-dimensional char density $(=\bar{\rho}_{char}/\rho^*)$ |
| $\rho_s$ | : | Non-dimensional density of unburnt solid fuel $(= \bar{\rho}_s/\rho^*)$ |
| $\rho_T$ | : | Non-dimensional total density of the fuel $(=\bar{\rho}_T/\rho^*)$ |
| $\bar{\rho}_{TO}$ | : | Initial total density $(=263\ kg/m^3)$ |
| $\rho_{TO}$ | : | Non-dimensional initial total density of the fresh fuel $(=\bar{\rho}_{TO}/\rho^*)$ |
| $\sigma$ | : | Stefan-Boltzmann constant $(5.678 \times 10^{-8}\ J/m^2/s/K^4)$ |

| $\xi, \mu, \eta$ | : | Direction cosines in *x, r* and *θ* directions |
| --- | --- | --- |
| $\varepsilon$ | : | Solid emittance |
| $\alpha$ | : | Solid absorptance |

**Subscripts**

| $B$ | : | Buoyant |
| --- | --- | --- |
| $b$ | : | Black body |
| $F$ | : | Fuel or flame |
| $g$ | : | Gas phase |
| $i$ | : | Species *i* |
| $R$ | : | Reference |
| $s$ | : | Solid phase |
| $w$ | : | Value at wall |
| $\infty$ | : | Value at far field |

**Superscripts**

| * | : | Evaluated at *T*\* |
| --- | --- | --- |
| - | : | Dimensional quantity |

# 1. Introduction

A thorough understanding of flame spread phenomena over solid combustible is important for fire safety, both on earth as well as in human space missions. In the process of flame spread, heat transfer from burning region to the unburned part of fuel plays an important role [1]–[3]. This heat transfer causes preheating and pyrolysis of the solid fuels. The unburnt combustible surface ahead of flame experiences elevated heat flux originating from not only flame but also hot neighbouring surfaces. So, the net heat transfer depends on the number of flaming surfaces and also on its distance from them. The heat transfer characteristics of spreading flame also depend on the orientation and geometry of the fuel [4]. The fuels can be like planar shapes [5]–[9], cylindrical [10]–[13], U shape [14] etc.. Apart from the geometry of the fuel, fuel orientation and multiple fuels arrangement also play a major role on flame spread [15]. There is an added complexity in the study of flame spread over combustible circular ducts as compared to a flat or solid cylindrical fuel due to presence of concave inner surface and convex outer surface. On the inner side of the circular duct, it can be seen that the flames can interact and influence each other along the circumference of the fuel (Fig. 1). In the enclosed region flow and oxygen availability may also vary. The outer surface, on the other hand, has flame, exposed directly to the ambient.. *Itoh et al* [16] performed experiments and theoretical analysis for the flame spread over circular ducts in normal gravity environments. The experiments were performed over hollow cylinders made of 0.2 mm thick paper and of diameters varying from 8 mm to 200 mm. From this study, it was concluded that flame spread rate increases with increase in fuel diameter. The increase of flame spread with fuel diameter was reported to be because of the increase in the radiative heat feedback coming from ember and flame at inner region of the fuel. While there has been a limited study on flame spread over thin circular duct geometry, valuable insights can be obtained from the studies of flame spread over thin parallel fuel configurations. A key similarity between the circular duct and thin parallel fuel configurations is the interacting flames at the inner side of the fuel. One of the earliest experimental and analytical work is that of *Kurosaki and Chiba* [17]. It was reported that for two parallel fuels, when the separation distance was small (<0.3 cm) the flame spread rate was reduced by half as compared to flame spread over a single sheet. It was also reported that a maximum flame spread rate was observed. when the separation distance was around 1.5 cm to 2 cm. This was explained to be due to increase in conduction heat transfer supplemented with radiation heat transfer. A further study on flame spread over multiple parallel thin sheet by Itoh and Kurosaki [18] showed that , as the number of thin fuel sheets increases the flame spread rate increases and flame spread rate remains constant for fuel sheets more than seven. *Shih and Hong* [4] experimentally studied upward flame spread over single solid, two and four solids in parallel, partially enclosed and fully enclosed configurations. The increase in flame spread over multiple solid fuels was reported to be because of the convection and radiation heat transfer from neighboring flames as well as due to flow channeling effects. A numerical study on opposed flow flame spread over parallel fuels was reported by *Malhotra et al*.[19]. In this study, fuel sheets in an infinite array configuration studied in microgravity environment to understand effect of spacing between sheets on flame spread rate. Both, flame spread rate and limiting oxygen index (LOI) showed a non-monotonic trend with spacing between the fuel sheets.  In a recent numerical and experimental work on flame spread over multiple parallel fuel sheets presented [20], it was reported that flame spread rate shows a non-monotonic trend with the spacing distance between fuels, and a peak flame spread rate was observed at 1 cm fuel spacing distance. This non-monotonic trend of flame spread rate with inter-fuel spacing was reported to be due to two competing effects. One is the increase in oxygen availability which enhances the flame spread rate and the second is the decrease in heat feedback from the flame to the unburnt fuel as the spacing distance increases. *T. Matsuoka et al.*[21] studied the flame spread inside two narrow channel namely parallel plate and cylindrical, made of PMMA. The effect of geometry and channel height on flame spread rate are observed. It was reported that flame spread rate for both the geometry follow non-monotonic increasing decreasing trend with respect to channel height and for larger value the channel height flame spread rate approaches to flat PMMA. The circular duct configuration, which is also different from parallel fuel configuration as the solid space is curved and inner region of fuel is fully bounded by fuel/solid surface has not been explored much.

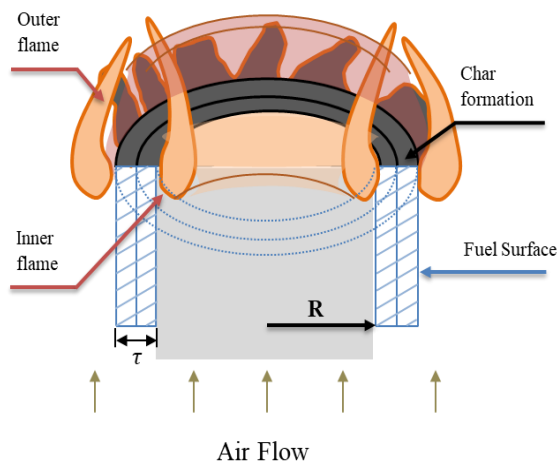

Fig.1 Model of flame spread over circular duct

In the present work, a numerical model is developed to study the flame spread over circular duct in normal gravity and microgravity environments. The solid fuel model also accounts for char in addition to pyrolysis model. The model is validated with inhouse experiments in both normal gravity and microgravity at varying fuel radius ranging from 5 mm to 24.5 mm, and flow speed ranging from quiescent to 30 cm/s. The effect of fuel radius and flow speed on flame spread rate is also studied. The experimental measurements and numerical simulation complement each other and helpful to gain valuable insights on flame spread over circular duct. The numerical model and experimental setup for flame spreading over circular duct is discussed.

## 2. Numerical Modelling

The pyrolysis and combustion process of flame spreading over hollow circular duct is modelled as a 2-D axisymmetric numerical model. The simplified assumptions include: laminar flow as velocities are small (< 1 m/s), all gas components follow the ideal gas equation and specific heats are functions of temperature only. The fuel is assumed to be thermally thin which implies there is no temperature gradient along the thickness of fuel. This condition holds true when thermal diffusion time is less compared to flame spread time scale. The fuel is assumed to compose of combustible cellulose and carbonaceous char.

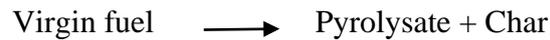

Virgin fuel ⟶ Pyrolysate + Char

A single-step, second-order finite-rate Arrhenius reaction between fuel vapor and oxygen is assumed. The specific heats are function of temperature for each species and all components of gases are assumed to follow ideal gas law.

The stoichiometric combustion of fuel pyrolysate in air can thus be written as:

$$C_6H_{10}O_5 + 6(O_2 + 3.76N_2) \rightarrow 6CO_2 + 5H_2O + 22.56N_2 \qquad (1)$$

For the above one-step cellulose and air stoichiometric reaction, the stoichiometric ratios are $(f_F = -1,\ f_{O2} = -1.1852,\ f_{co2} = 1.6296,\ f_{H2O} = 0.5556,\ f_{N2} = 3.901$

The present 2D axisymmetric code for circular duct fuel is developed from an existing 2D axisymmetric model used for study of flame spread over thin solid cylindrical fuel [11] which has been used successfully to predict various aspect of flame spread over thin fuels. The governing equation and boundary conditions are described in the next sections.

### 2.1. Gas Phase Model

The gas phase equations consist of non-dimensional form of governing equations of Navier-Stokes for laminar flow, along with conservation equations of mass, energy and species in cylindrical coordinates. The species equations are written for fuel vapor, oxygen, carbon dioxide and water vapours. The reaction between oxygen and fuel vapour is assumed as one-
9

step, second-order finite-rate Arrhenius kinetics. The gas phase conservation equations are similar as summarized in previous article [12].

### 2.1.1. Continuity equation

$$\frac{\partial \rho}{\partial t} + \frac{\partial \rho u}{\partial x} + \frac{\partial \rho v}{\partial r} + \frac{\rho v}{r} = 0 \tag{2}$$

### 2.1.2. Axial momentum equation

$$\frac{\partial(\rho u)}{\partial t} + \nabla\left[\rho u \boldsymbol{V} - \frac{\mu}{Re}\nabla u\right] = -\frac{\partial p}{\partial x} + \frac{1}{Re}\left[\frac{\partial}{\partial x}\left(\frac{1}{3}\mu\frac{\partial u}{\partial x} - \frac{2\mu}{3}\frac{\partial v}{\partial r}\right)\right] + \frac{1}{Re}\frac{\partial}{\partial r}\left(\mu\frac{\partial v}{\partial x}\right)$$

$$+ \frac{1}{Re}\left[\frac{1}{r}\left(\frac{\mu}{3}\frac{\partial v}{\partial x}\right)\right] + \left(\frac{U_B}{U_R}\right)^3 \frac{(\rho - \rho_\infty)}{(\rho_s - \rho_\infty)} g - \left(\frac{dv_f}{dt}\right)\frac{(\rho - \rho_\infty)}{(\rho_f - \rho_\infty)} \tag{3}$$

### 2.1.3. Radial momentum equation

$$\frac{\partial(\rho v)}{\partial t} + \nabla\left[\rho v \vec{V} - \frac{\mu}{Re}\nabla v\right] = -\frac{\partial p}{\partial r} + \frac{1}{Re}\left[\frac{\partial}{\partial x}\left(\mu\frac{\partial u}{\partial r}\right)\right] + \frac{1}{Re}\left[\frac{\partial}{\partial r}\left(\frac{\mu}{3}\frac{\partial v}{\partial r} - \frac{2}{3}\mu\frac{\partial u}{\partial x}\right)\right]$$

$$+ \left[\frac{1}{r}\left(\frac{1}{3}\mu\frac{\partial v}{\partial r} - \frac{4}{3}\mu\frac{v}{r}\right)\right] \tag{4}$$

here

$$v_f = \text{non} - \text{dimensinal flame spread rate} = \frac{V_f}{\overline{U}_R}$$

$V_f$ = dimensional flame spread rate     $\overline{U}_R$ = reference velocity

$Re$ = Reynolds number ($=\rho^*\overline{U}_R\overline{L}_R/\mu^*$)

### 2.1.4. Specie equation

$$\frac{\partial(\rho Y_i)}{\partial t} + \frac{\partial}{\partial x}\left(\rho Y_i u - \left(\frac{1}{Le_i}\right)D_i\frac{\partial(\rho Y_i)}{\partial x}\right) + \frac{\partial}{\partial r}\left(\rho Y_i v - \left(\frac{1}{Le_i}\right)D_i\frac{\partial(\rho Y_i)}{\partial r}\right)$$

$$= \dot{\omega}_i''' + \left(\frac{1}{Le_i}\right)\frac{1}{r}D_i\frac{\partial(\rho Y_i)}{\partial r} - \frac{\rho v Y_i}{r} \tag{5}$$

i = F, O₂, CO₂, H₂O, N₂     $Y_i$ = mass fraction of species i

$D_i$ = Diffusion coefficient of species $i$

$\dot{\omega}_i'''$ = Sink/source term in gas-phase reaction = $f_i\dot{\omega}_F''' = f_i Da\rho^2 Y_F Y_{O2} \exp(-E_g/T))$

$f_i$ is the stoichiometric mass ratio of species i and fuel

$$Da = \text{Damköhler nember} = \frac{\text{characteristic flow time}}{\text{characteristic reaction time}}$$

$$= \frac{\bar{L}_R/\bar{U}_R}{\rho^*/\rho^{*2}\bar{B}_g} = \frac{\bar{B}_g \rho^* \bar{L}_R}{\bar{U}_R} = \frac{\alpha^* \rho^* \bar{B}_g}{\bar{U}_R^2}$$

Here, $\bar{L}_R$ is reference length, $\alpha^*, \rho^*$ are thermal diffusivity and density of the gas phase at reference temperature $T^*(1250\ K)$ respectively and $\bar{B}_g$ is gas phase pre-exponential factor.

$$Le_i = \text{Lewis number of species i} = \frac{\text{diffusion time}}{\text{convection/conduction}} = \frac{\overline{L_R^2/D_i^*}}{\bar{L}_R/\bar{U}_R} = \frac{\overline{L_R^2/D_i^*}}{L_R^2/\alpha^*} = \frac{\alpha^*}{D_i^*}$$

($Le_F = 1, Le_{O2} = 1.11, Le_{CO2} = 1.39, Le_{H2O} = 0.83, Le_{N2} = 1$)

### 2.1.5. Energy equation

$$\frac{\partial(\rho c_p T)}{\partial t} + \frac{\partial}{\partial x}\left(\rho c_p u T - k\frac{\partial T}{\partial x}\right) + \frac{\partial}{\partial r}\left(\rho c_p v T - k\frac{\partial T}{\partial r}\right)$$

$$= \sum_{i=1}^{N}\left(\frac{1}{Le_i}\right)\rho D_i c_{pi} \vec{\nabla} Y_i \vec{\nabla} T - \sum_{i=1}^{N} h_i \dot{\omega}_i''' + \frac{1}{r}k\frac{\partial T}{\partial r} + \vec{V} c_p \vec{\nabla} T \left(\frac{k}{c_p}\right) - \left(\frac{1}{B_0}\right)\vec{\nabla}.\vec{q}_r \qquad (6)$$

$c_{pi}$ is non-dimensional specific of species i

$c_p$ is non-dimensional gas- phase specific heat and it can be calculated as $\sum_{i=1}^{N} c_{pi} Y_i$

$k$ is non-dimensional gas phase thermal conductivity

$h_i$ is non-dimensional enthalpy of species $i = \left(\bar{h}_i^o + \int_{\bar{T}_o=298\ K}^{\bar{T}} \bar{c}_{p,i} d\bar{T}\right)/c_p^* \bar{T}_\infty$

$$B_0 = \text{Boltzman number} = \frac{\rho^* c_p^* \bar{U}_R}{\sigma \bar{T}_\infty^3}$$

The term $\nabla.\vec{q}_r$ is the divergence of radiation heat flux which accounts for energy exchange due to presence of radiatively participating media

### 2.2. Gas Phase Boundary Conditions

The governing equations are of elliptical nature and require boundary conditions to be specified at all boundaries. These boundary conditions are listed over the boundaries as shown in computational domain in the Fig. 2.

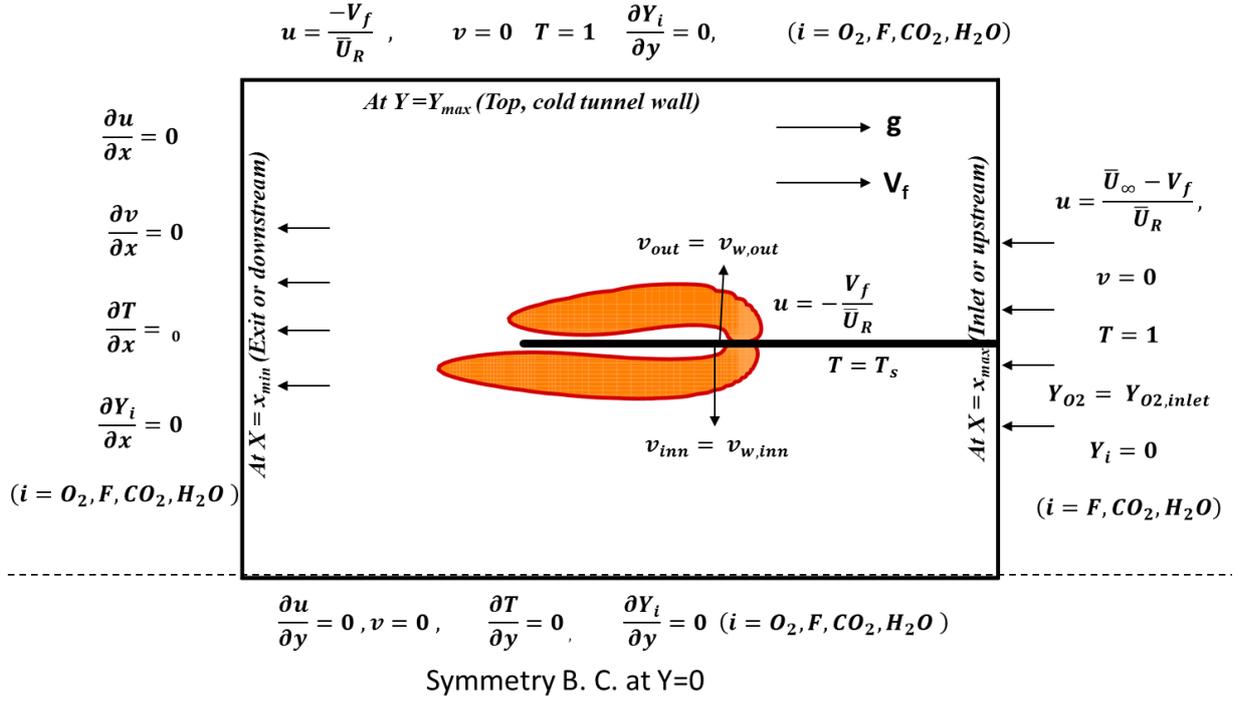

Fig.2. Representation of computational domain and gas phase boundary conditions

## 2.3. Solid Phase Model

The solid phase model for hollow cylindrical fuel consists of mass and energy conservation equations with a first order pyrolysis law for the pyrolysis of virgin fuel. As mentioned before, the solid phase is assumed as thermally thin so the conservation equations are solved in only one direction.

### 2.3.1. Mass conservation equation

$$\frac{\partial \rho_T}{\partial t} - v_f \frac{\partial \rho_T}{\partial x} = -\left(Y_{pyro}\dot{m}'''_s + \dot{m}'''_{char}\right) \tag{7}$$

### 2.3.2. Energy conservation equation

$$\begin{aligned}
\frac{\partial \rho_T c_s T_s}{\partial t} dV - v_f A_{CS} \frac{\partial \rho_T c_s T_s}{\partial x} dx \\
= \left(q''_{c,i} A_{Surf,i} + q''_{c,o} A_{Surf,o}\right) + \frac{1}{Bo}\left(q''_{r,i} A_{Surf,i} + q''_{r,o} A_{Surf,o}\right) \\
- \dot{m}'''_s c_s \left((1-c)T_l - L + T_s(c - Y_{pyro})\right) dV + \dot{m}'''_{char} c_s T_s dV
\end{aligned} \tag{8}$$

Here

$A_{CS}$ = Cross-sectional area of the hollow fuel

$q''_{c,i}$ = Conduction heat flux from gas phase at the inner surface of the fuel.

$q''_{c,o}$ = Conduction heat flux from gas phase at the outer surface of the fuel

$q''_{r,i}$ = Net radiation heat flux at the inner surface of the fuel.

$q''_{r,o}$ = Net radiation heat flux at the outer surface of the fuel

$A_{Surf,i}$ = Inner surface area of the fuel

$A_{Surf,o}$ = Outer surface area of the fuel

In the above equations, eqn. 2 shows conservation of mass and in equation 3, the first term on the RHS is the combined conduction heat from the gas phase to the solid at the inner and outer surface of the fuel, second term on the RHS denotes the combined net radiation at the inner and outer surface, third term is the heat of pyrolysis and the last term is the heat absorbed because of char.

### 2.4. Solid Phase Boundary Conditions

The boundary conditions for the solid phase governing equations are the prescribed solid fuel density and the surface temperature at the fuel leading edge upstream of the flame at $(x = x_0)$.

$$T_s = 1, \ \rho_T = \rho_s = \bar{\rho}_s / \rho^*$$

## 3. Radiation Transport

The radiation heat flux and its divergence appear in solid and gas phase governing equations and needs to be determined. Radiative heat flux and its divergence are calculated from the radiation intensity for which the Radiative Transfer Equation (RTE) is solved. The RTE for radiation energy passing in a specified direction $\vec{\Omega}$ is presented below

$$(\vec{\Omega}.\nabla)I(\vec{r},\vec{\Omega}) = k(\vec{r})I_b(\vec{r}) - \beta(\vec{r})I(\vec{r},\vec{\Omega}) + \frac{\sigma_s(\vec{r})}{4\pi}\int_{\Omega'}\Phi(\vec{\Omega}',\vec{\Omega})I(\vec{r},\vec{\Omega}')d\vec{\Omega}' \qquad (9)$$

$I$ = radiation intensity

$I_b$ = blackbody intensity at local temperature, $\beta = k+\sigma_s$, is the extinction coefficient

$k$ = absorption coefficient

$\sigma_s$ = scattering coefficient

$\Phi$ = scattering phase function

$\vec{\Omega}$ = unit vector specifying the direction of scattering through a position $\vec{r}$

$\vec{r}$ = position vector

Since solving Radiative Transfer Equation with spectral accuracy is computationally prohibitive for this coupled multi-dimensional problem, it is solved by Discrete Ordinates Method (DOM). DOM is based on a discrete representation of the directional variation of the radiative intensity, further details of DOM can be found in literature [22]

The transfer equation for radiation energy passing in a specified direction $\vec{\Omega}$ through a small differential volume in an emitting, absorbing and non-scattering gray medium, in two dimensional co-ordinates can be written in S-N approximation as

$$\xi \frac{\partial I(x,r,\vec{\Omega})}{\partial x} + \frac{\mu}{r}\frac{\partial I(x,r,\vec{\Omega})}{\partial r} - \frac{1}{r}\frac{\partial [\eta I(x,r,\vec{\Omega})]}{\partial \psi} + \beta I(x,r,\vec{\Omega}) = S(x,r,\vec{\Omega}) \qquad (10)$$

Where $\xi, \mu$ are the direction cosines of $\vec{\Omega}$ and are defined as

$$\xi = cos\theta, \mu = sin\theta cos\psi$$

$\theta$ is the polar angle and $\psi$ is the azimuthal angle measured from the r direction. Source term $S(x,r,\vec{\Omega})$ is given by

$$S(x,r,\vec{\Omega}) = k(x,r)I_b(x,r) + \frac{\sigma_s(x,r)}{4\pi}\int_{\Omega'} \Phi(\vec{\Omega}',\vec{\Omega})I(x,r,\vec{\Omega}')d\vec{\Omega}' \qquad (11)$$

$\beta = k$ (Assuming the medium to be non-scattering)

$$S(x,r,\vec{\Omega}) = k(x,r)I_b(x,r)$$

The boundary condition for the outgoing radiation intensity from the outer fuel surface can be expressed as

$$I(x,r,\vec{\Omega}) = \varepsilon I_b(x,r) + \frac{1-\alpha}{\pi}\int_{n\Omega'<0} |n.\Omega'|I(x,r,\Omega')d\Omega' \qquad n.\Omega > 0 \qquad (12)$$

And boundary condition for radiation intensity at inner surface can be expressed as

$$I(x,r,\vec{\Omega}) = \varepsilon I_b(x,r) + \frac{1-\alpha}{\pi}\int_{n\Omega'>0} |n.\Omega'|I(x,r,\Omega')d\Omega' \qquad n.\Omega < 0 \qquad (13)$$

$\varepsilon, \alpha$ are emissivity and absorptivity of the wall surface.

Once the radiation intensity field is obtained, the total incident radiation, radiation heat fluxes in $x$ and $r$ directions, and the divergence of the heat flux are obtained as follows:

The incident radiation can be written as

$$G(x,r) = \int_{4\pi} I(x,r,\vec{\Omega})d\vec{\Omega} \qquad (14)$$

The radiation heat fluxes in $x$ and $r$ directions are respectively

$$q_r^{x+}(x,r) = \int_{\xi>0} \xi I(x,r,\vec{\Omega})d\vec{\Omega}$$

$$q_r^{x-}(x,r) = \int_{\xi<0} \xi I(x,r,\vec{\Omega})d\vec{\Omega}$$

$$q_r^{r+}(x,r) = \int_{\mu>0} \mu I(x,r,\vec{\Omega})d\vec{\Omega}$$

$$q_r^{r-}(x,r) = \int_{\mu<0} \mu I(x,r,\vec{\Omega})d\vec{\Omega}$$

Net radiation heat flux in $x$ and $r$ directions are calculated as

$$q_r^x(x,r) = q_r^{x-}(x,r) + q_r^{x+}(x,r)$$

$$q_r^r(x,r) = q_r^{r-}(x,r) + q_r^{r+}(x,r)$$

For the assumed emitting, absorbing and non-scattering gray medium, the divergence of the radiative flux is shown below.

$$\nabla \cdot \vec{q}_r = k(x,r)[4\sigma T^4(x,r) - G(x,r)]\text{nb}$$

## 4. Solution Procedures

The partial differential equations of elliptical nature for gas phase and flow problem are solved numerically by the SIMPLER algorithm (Patankar, 1980). The non-linear gas-phase equations are discretized using a finite-volume technique. The velocities are stored at staggered grid locations with respect to scalar variables. The resulting set of algebraic equations is solved by sweeping line-by-line in each direction where a combination of Gauss-Seidel and the tri-diagonal matrix algorithm (TDMA) is used. In addition, the gas-phase system is coupled to the solid-phase equations. The solid-phase equations are solved by the finite-difference technique. The steady flame spread rate (the eigenvalue of the whole system) is determined iteratively by using a bisection method to force and the pyrolysis front (95% of the fresh fuel density) to occur at X = 0.

Since the radiation equation and the rest of the combustion/fluid equations are coupled, they are solved iteratively. In the present work the radiation routine was invoked for every 20 gas-solid iterations. Convergence to a steady state solution is determined by fixed flame spread rate over a couple of thousand iterations in conjunction with satisfaction of continuity to the order less than $10^{-3}$. Apart from convergence of continuity equation, the convergence of other governing equations is ensured by prescribing convergence for residue to be below $10^{-4}$ following the $L_\infty$ norm.

The present numerical study is carried out with non-uniform mesh to appropriately resolve the reaction zone in the flame leading edge, which is critical for predicting the flame spread phenomena. The grids are clustered near the flame leading edge or the flame anchor location to capture accurately the steep gradients. The grid expands in x and r-directions further away from the flame leading edge and fuel surface respectively. The governing equations are non-dimensional and the grid independence study has been carried out to determine the minimum grid size in both x and r-directions. The details of the mesh size at the flame anchor location in x and r directions are given in the table below. The minimum grid size of 0.05×0.005 is chosen for all calculations since further grid refinement has a negligible effect on the flame spread rate.

| Total number of grids in x and r-direction | Minimum grid size in x-direction ($\times \overline{L_R}$) | Minimum grid size in r-direction ($\times \overline{L_R}$) | Flame spread rate (mm/s) |
|---|---|---|---|
| 318×207 | 1.00 | 0.100 | 12.2 |
| 334×227 | 0.10 | 0.010 | 9.8 |
| 335×229 | 0.05 | 0.005 | 9.50 |
| 339×236 | 0.03 | 0.003 | 9.49 |

Table-1: Flame spread rates at different grid sizes in radial and axial directions.

## 5. Experiments

The experiments are performed using the same setup in normal gravity and microgravity environment. For microgravity tests, a 2.5 s drop tower facility available at the National Centre for Combustion Research and Development (NCCRD), IIT Madras, India, is used. The details on the design and operation of this drop tower facility can be found in the literature[23]. Fig.3 presents an experimental module used in this study. The circular duct fuel is placed at the centre of a 12 cm ×12 cm inner cross section and 53 cm long rectangular duct (Fig. 3 (a)). The cross-sectional view of the rectangular duct is shown in Fig. 3 (b) and it consists of two sections, an upper and a lower section. The lower section with 28 cm length consists of a DC fan, honeycomb, and a fine mesh while the fuel and ignite are housed in the upper section. A small DC fan is placed at the bottom of the duct which is capable of providing varying opposed flow speed. Honeycomb and fine mesh are placed in-between fuel holder and fan to obtain a uniform desired opposed flow in the test section. The upper section of the duct is made of transparent acrylic (with length equal to 25 cm) to observe the combustion phenomena over the solid fuel. The fuel sample is held at centre of the rectangular duct with the help of different thin metallic fuel holders as shown in Fig. 3(c). The whole experimental module is fixed over the top deck of the inner capsule, whereas the bottom two deck of the inner capsule has provision for data acquisition modules and power supply unit (Fig. 3(d)).

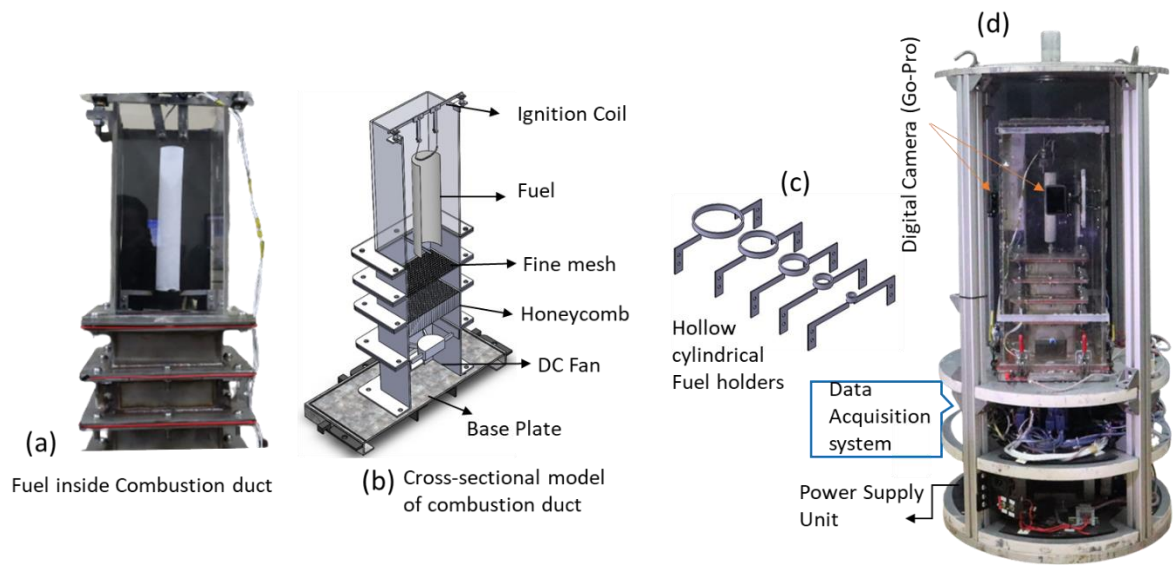

Fig. 3. The experimental module (a) Sectional view of combustion duct (b) Experimental set up enclosed with polycarbonate enclosure (c) Fuel sample holders (d) Complete experimental module placed over inner capsule.

Because of the limited availability of time (2.5 s) available for microgravity environment using drop tower facility, it is better to select a fuel in such a way that flame spreads rapidly and quickly achieve a steady state condition. The fuel used in the present experiment studies is 76-µm thick cellulosic paper (Kim-wipes) with a composition of 99 % cellulose and 1% polyamide. The area density of this paper is about 18 gsm (grams per square meter). The height of fuel sample is fixed at 180 mm which is fitted inside the combustion duct. The radius of the circular duct varied from 5 mm to 24.5 mm. The maximum limit of radius (24.5 mm) is selected to ensure that the shape and cross-section of the hollow cylinder is not disturbed while igniting with ignition coil and during the flame spread process. For the preparation of the circular duct fuel samples, paper is wrapped around a metallic tube of desired diameter and both the edges of the paper are glued at few points using small drop of glue and then the metallic tube is removed by directly sliding over. The hollow fuels stand over 1.5 mm thick stainless steel fuel holders (fig. 3(c)).

The fuel samples are ignited in normal gravity and once the fuel is ignited the ignition coil power is switched off and then the capsule housing the experimental module is dropped for 2.5 s to obtain flame spread data in microgravity environments. The flame spread behaviour of different fuel specimens is recorded with the help of two digital cameras (Go-Pro Hero-10) from two perpendicular sides (Fig. 3 (d)). The cameras are set at 120 FPS and 2.7 k resolution (2704 pixel wide by 1520 pixel high) to better capture the images of spreading flames. A white LED light is provided in the background to uniformly illuminate the fuel samples. The purpose of illuminating the fuel is to avoid the reflections of light coming from the flame to the white surface of fuel and to identify the clear leading edge of spreading flames on fuel surface.

The data obtained from the camera is post processed using an in-house code written in MATLAB. A method of tracking the flame leading edge over the solid surface is employed which is then used to calculate the flame spread rate. To make the distinction in the intensity of the flame and the fuel surface, the uniformly illuminated, intensity which is of equal or higher than that of the white part of the fuel is cropped from all the frames using MATLAB crop function. The cropped image is converted into grey image and a threshold intensity of 100 is selected which corresponds to pyrolysis front. The position of this threshold intensity (corresponding to flame leading edge) is tracked along the length of the fuel in each frame and the data for the position of flame leading edge with time is obtained. A typical data corresponding to flame spread over 5 mm and 14.5 mm and 24.5 mm radius circular duct fuels are shown in Fig. 4. The average flame spread rate is taken as the slope of the position vs time plot. One can note that the variation of flame position with time is linear and the fuel radius 24.5 mm has higher slope as compared to 14.5 mm and 5 mm fuel radius which correspond to flame spread rates of 12.5 mm/s, 9.4 mm/s and 6.9 mm/s respectively.

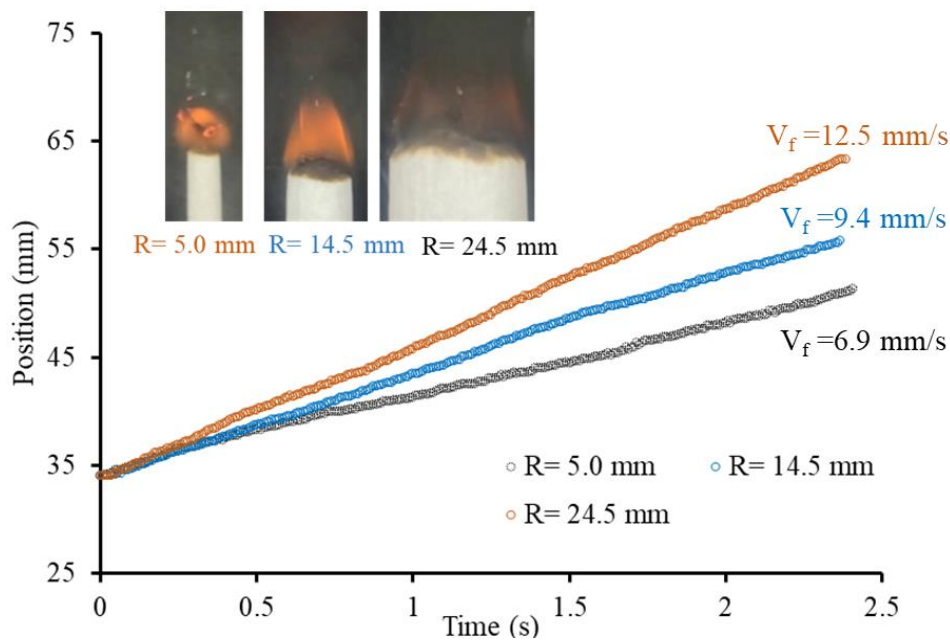

Fig.4. Position of flame leading edge versus time plot for circular duct having 5 mm and 14.5 mm and 25.5 mm fuel radius at 20 cm/s flow speed.

## 6. Results and Discussions

Numerical modelling for flame spread over thin circular ducts have been carried out in normal gravity and microgravity environments. To validate the numerical model experiments are also carried out in similar conditions. The results are explained simultaneously both in normal gravity and microgravity environments. In first part we explained about the flame shapes and temperature contours for different fuel radii and at different opposed flow speed in normal gravity and microgravity environments. While in later sections, the controlling mechanisms of heat and mass transport between the flame and solid fuels along with the flame spread rates at different conditions are discussed.

### 6.1. Spreading flames in normal gravity and microgravity

A comparison between numerical calculations and experimental observations of flame spreading over circular duct against an opposed flow velocity of 10 cm/s is shown in Fig. 5. Numerical calculations are done for three different radii (14.5 mm, 19 mm and 24.5 mm) while experiments are performed for five different radii (5 mm, 9.5 mm, 14.5 mm, 19 mm and 24.5 mm). Fig.5 (a) shows that the length of the flame increases as the radius of the fuel increases in both experiments and numerical simulations. Another distinct observation which can be made from the numerical results is the longer flame length at the inner side of the fuel as compared to flame at the outer side for each fuel radius. The longer flame at inner surface is due to oxidiser deficiency at lower fuel radii.

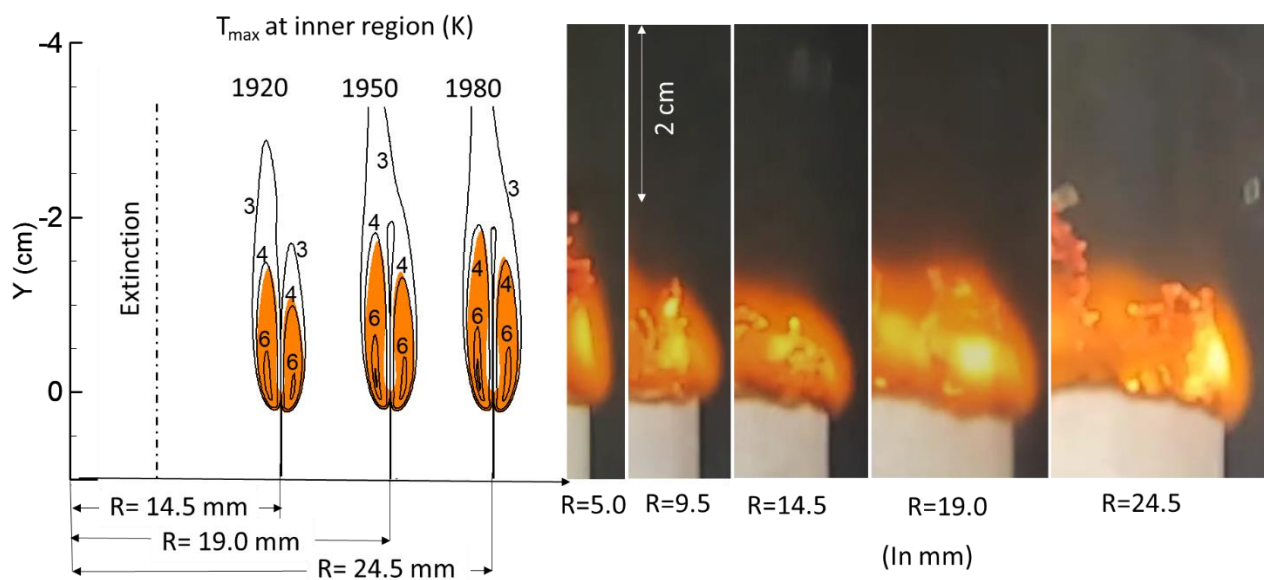

Fig.5. Spreading flames over different fuel radius in normal gravity environment at 25 cm/s flow speed (a) numerical (b) experimental

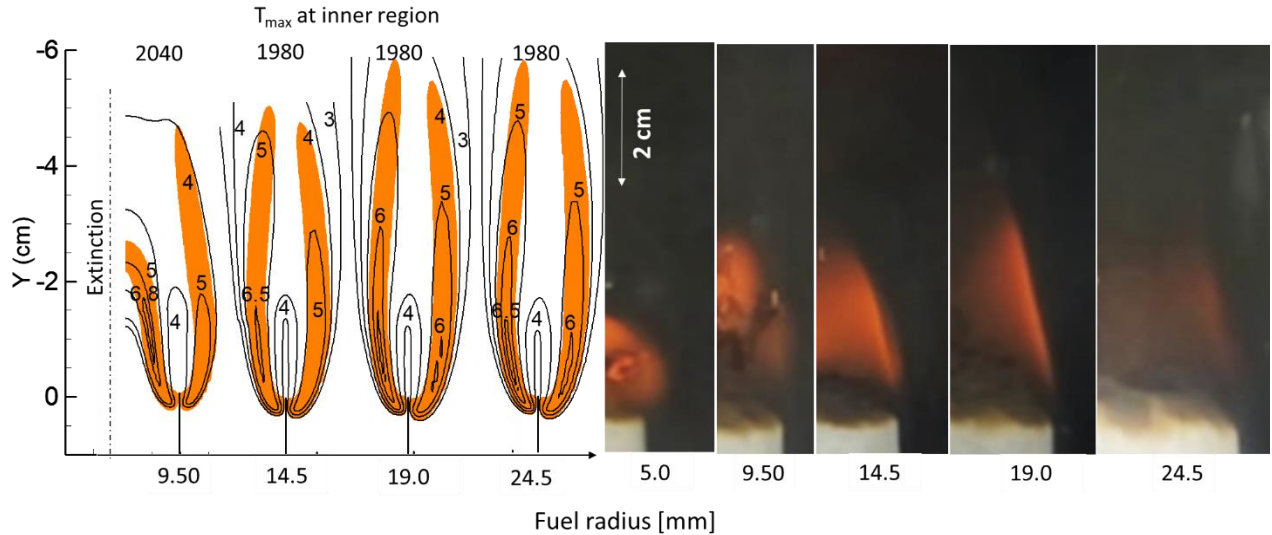

Fig.6. Spreading flames over different fuel radius at 25 cm/s flow speeds in microgravity gravity (a) numerical (b) experimental

Furthermore, a slight increase in maximum flame temperature at inner region is observed as the fuel radius is increased. The maximum flame temperature at inner region increases from 6.61 (=1983 K) for fuel radius 14.5 to 6.68 (2004 K) for fuel radius 24.5. If the fuel radius is increased further, one may see a decrease in flame temperature because of equal amount of oxidiser available at inner core region and outer region, which is not explored in the present study and can be investigated in future study. The increase in maximum flame temperature indicates that the flame becomes stronger with the increase in fuel radius considered in this study. A stronger flame enhances heat feedback from the burning zone to the unburnt solid fuel which eventually results in faster flame spread rate.

Fig.6. shows a comparison between numerical calculations and experimental observations of spreading flames over circular duct against an opposed flow velocity of 25 cm/s in microgravity environment. Numerical simulations are done for circular ducts with four different fuel radii (9.5 mm, 14.5 mm, 19 mm and 24.5 mm) while experiments are performed for five different radii (5 mm, 9.5 mm, 14.5 mm, 19 mm and 24.5 mm). Similar to normal gravity, here also, It the length of the flame increases as the radius of the fuel increases in both experiments and numerical simulations (Fig. 6). As the radius of fuel increases from 9.50 mm to 24.5 mm, the flame length increases from 5 cm to 6.1 cm in the numerical simulation (Fig. 6 (a)). Due to illumination of LED light in the experiments to track the pyrolysis front and to calculate the flame spread rate, the actual flame size is not visible however in general increase in flame length with increase in fuel radius is also observed in experiment.

Another distinct observation which can be made from the numerical results is that for fuel radius of 9.50 mm the flames are merged at inner core region and maximum flame temperature is higher (6.8=2040 K) as compared to other fuel radii (6.6= 1980 K). Further increase in fuel radius, flame length at the inner side of the fuel is slightly higher as compared to flame at the outer side for each fuel radius. The longer flame at inner surface is due to oxidiser deficiency at lower radius fuel. As the radius of the fuel increases the mass of the fuel coming out from fuel surface ($\dot{m}_f = 2\pi R \tau \rho_f V_f$) also increases and length of flame increases. The flame length at outer surface for 9.50 mm fuel radius is about 4.7 cm long, and increases to 5.8 cm for fuel radius of 24.5mm. In microgravity, the flame characteristic such as flame length and luminosity are seen to be more affected with respect opposed flow speed.

Figure 7 show these flame shapes at opposed flow speed of 0 cm/s, 10 cm/s, 20 cm/s and 30 cm/s in normal gravity environments obtained from numerical calculations and experiments respectively. The presented flames are spreading over the 14.5 mm fuel radius and one can see that as the flow speed increases in normal gravity the flame length first increases in the flow speed range of 0 cm/s to 10 cm/s then further increase in flow speed from 10 cm/s to 30 cm/s, the flame length decreases.

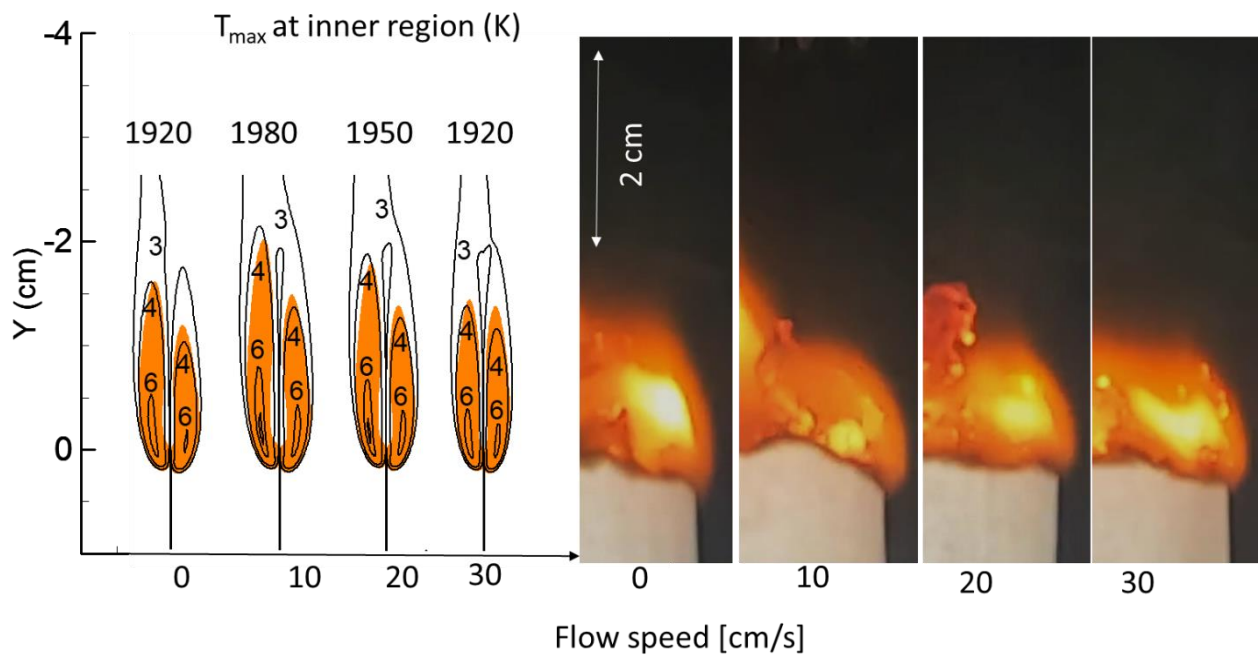

Fig.7. Spreading flames over 14.5 mm fuel radius at different flow speeds in normal gravity (a) numerical (b) experimental

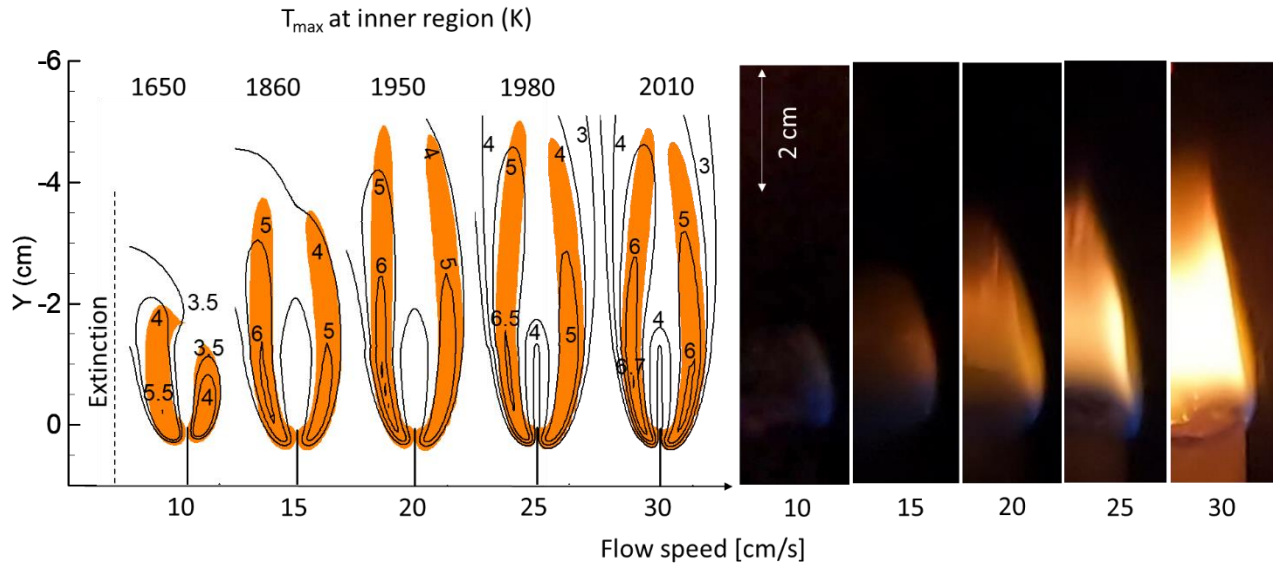

Fig.8. Spreading flames over 14.5 mm fuel radius at different flow speeds in microgravity (a) experimental (b) numerical

The maximum temperature at inner flame also first increase from 6.4 (1920 K) to 6.6 (1980 K) when flow speed increase from 0 cm/s to 10 cm/s then further increase in flow speed from 10 cm/s to 30 cm/s the maximum temperature decreases at inner core region and reduces from 6.6 (1980 K) to 6.4 (1920 K) while at outer flame surface, the maximum flame temperature decreases from 6.2 (1860 K) to 6.1 (1830 K) as the flow speed is increased from 10 cm/s to 30 cm/s. The increase of flame tempearture and flame length in velocity range 0 cm/s to 10 cm/s is due to increased availability of oxygen at inner core region of fuel. Further the reduction of flame temperature with increase in flow speed in range of 10 cm/s to 30 cm/s, is due to the convective cooling effect of the flame in normal gravity environment. Furthermore in numerical calucations the flames at inner and outer surface is asymmetric at 10 cm/s, but as the flow speed is increases the flame at inner core region decrease from 2.2 cm flame length to 1.7 cm length and become almost equal in length with that of the flame at the outer region at 30 cm/s flow speed.The non monotonic increasing decreasing trend of flame length and flame temperature will affect the flame spread rate trend with respect to opposed flow speed which is discussed in the later section.

Figure 8 show the spreading flame shapes at opposed flow speed of 10 cm/s to 30 cm/s in increasing steps of 5 cm/s in microgravity environments obtained fom numerical calculations and experiments respectively. The presented flame are spreading over the 14.5 mm radius and one can see that as the flow speed increases in microgravity, the luminosity of the flame increases. The increase in luminosity of the flame with increase in flow speed is due to increase in the flame temperature. In the presented numerical calculations the maximum temperature at inner core region increases from 5.5 (1650 K), for 10 cm/s flow speed to 6.7 (2010 K) for 30 cm/s flow speed. Whereas the maximum flame temperature at outer surface at 10 cm/s flow speed, increase from 4.9 (1470K) to 6.2 (1860 K) at 30 cm/s flow speed.

### 6.2. Surface temperature, fuel vapor velocity and normaized density of circular duct in normal gravity and microgravity

Fig. 9 (a) shows the variation of solid surface temperature (right axis) and fuel vapor velocity (left axis) for different fuel radius over the fuel length near flame anchor location in normal gravity. Figures in insets show enlarge plots of temprature and fuel vapor velocity in the preheat region. The solid temperature rises rapidly from ambient temperature to pyrolysis temperature (720 K) where the solid fuel starts to pyrolyse in normal gravity. Further downstream, the solid temperature increases and flattens. In the downstreams locations where, all the fuel vapor gets consumed by the flame, the temperature of the char remains high (950 K) which eventually heats the unburnt solid in the preheat region at inner core region by radiation via gas. It can be seen from the figure (inset of Fig. 9(a)), that the solid temperture is higher for 24.5 mm fuel radius as compared to 19 mm and 14.5 mm fuel radius resulting, in faster pyrolysis ( higher mass loss rate or higher fuel vapor velocity) for fuel of lager radius and hence prediction of higher flame spread rate.Similar to Fig.9 (a), in the Fig. 9 (b) the left axis shows the solid surface temperature and right axis show the solid fuel density and char density ( densities are normalized with unburnt virgin fuel density) in normal gravity. It can be seen from the figure that the solid fuel temperature starts to increase from ambient to about 900 K at the locations near flame anchor location (0.5 cm to -0.18 cm). At the location where temperature is sufficently high for the fuel to pyrolyse, one can note that the decrease in solid fuel density (Fig. 9 (b)). For all the fuel radius considered here, pyrolysis of the fuel is completed at a downstream location of about -0.18 cm (inset Fig. 9 (a)) which is also reflected in Fig. 9 (b), where the solid fuel density decreases to zero. The solid fuel is assumed to consist of pyrolysate and char so, as the solid fuel degrades, a part of it produces pyrolysate and the remaining is left as char. In Fig. 9 (b), it can be seen that the char density increases from zero at far upstream region to 15% towards the downstream of flame anchor location (X=0). The amount of char present in the fuel is taken 15% of the unburnt fuel density in the numerical nodel from TGA data.

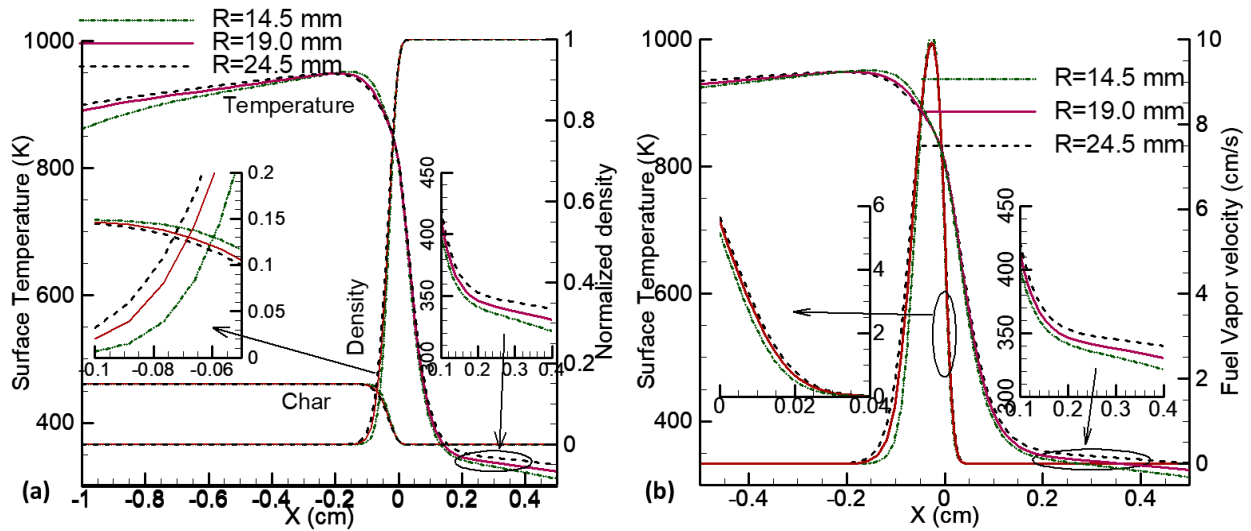

Fig.9. Surface temperature and fuel vapour velocity (a) and normalized density of virgin fuel and char (b) for different fuel radius in normal gravity.

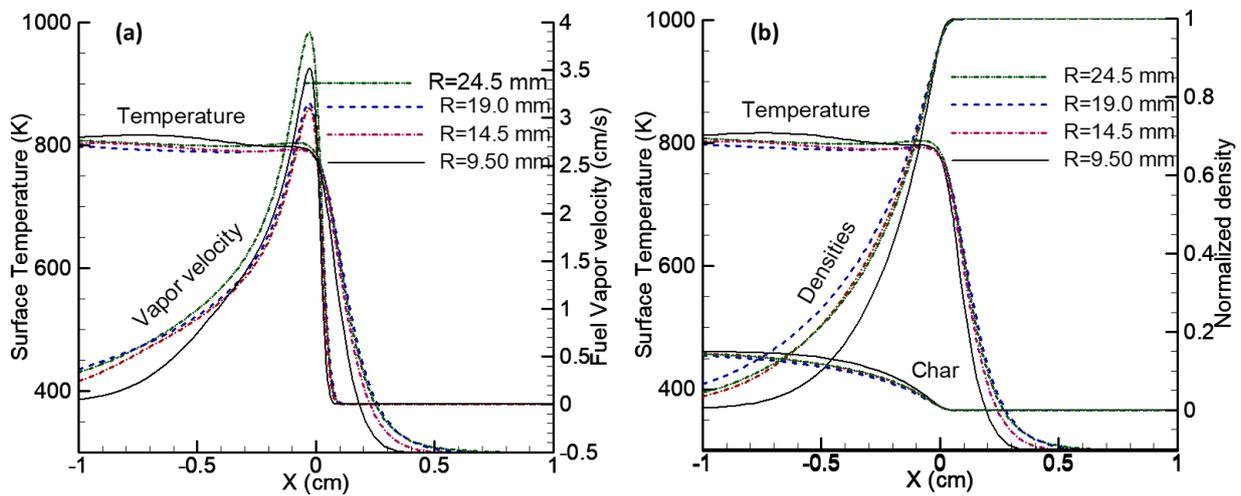

Fig. 10. Surface temperature and fuel vapour velocity for different fuel radius in microgravity at 25 cm/s flow speed in microgravity

In microgravity the variation of solid surface temperature (left axis) and fuel vapor velocity (right axis) for different fuel radius along the fuel length near flame anchor location in microgravity is shown in Fig. 10 (a). The maximum fuel vapor velocities in microgravity is less (~3 cm/s to 4 cm/s) as compared to normal gravity (~10 cm/s in Fig. 9) for different fuel radii. In the downsream when all the fuel vapor gets consumed by the flame in microgravity, the temperature of the char remains high (800 K) but lower than char temperature in normal gravity (950K), which eventually heats the unburnt solid in preheat region at inner core region by radiation. It can be seen from the figure that the solid temperture is higher for 24.5 mm fuel radius in preheat region followed by 19 mm, 14.5 mm and 9.5 mm fuel radius resulting, the fuel vapor velocity of different radius fuels follows the solid temperature trend in microgravity.

In Fig. 10 (b) the left axis shows the solid surface temperature and right axis show the solid fuel density and char density ( densities are normalized with unburnt virgin fuel density). One can note that, the solid temperature starts to increase from ambient to about 800 K. as the temperature rises the pyrolysis of fuel progresses, the fuel density starts decreasing. The pyrolysis length in normal gravity (0.18 cm) increases to about 1 cm in microgravity. The char density increases from zero at far upstram region to 15 % at location further downstream of flame anchor location (X=0) as compared to normal gravity.

### 6.3. Heat flux distribution over fuel surface in normal gravity and microgravity

Figure 11 presents heat flux distribution at outer and inner surface of the circular duct along with flame. This heat flux distribution correspond to radius of 14.5 mm at 25 cm/s flow speed. The conduction heat flux at outer ($q_{c,o}$) and inner ($q_{c,i}$) surface increases from preheat region and peaks at location near the flame achor location (X=0). Also in Fig. 8 the radiation heat fluxes curve shows radiation heat flux at inner and outer surface of solid fuel. At the outer surface, the radiation heat flux shown, is a combination of radiation heat loss from the fuel surface and radiation heat gain from the hot gas. While in the case of inner surface, hot char also contributes in heating up of fuel through radiation. One can note that the radiation heat flux at outer surface ($q_{r,o}$) shows small magnitude (~0) in preheat region and decreases to negative values towards downstream locations however, radiation heat flux at inner section ($q_{r,i}$) has positive magnitude along preheat region indicating heat gain because of radiation. The integrated values of these heat fluxes per unit circumference are also shown in table in next section. The summation of coduction heat flux and net radiation, shown as the net heat flux at outer ($q_{net,o}$) and inner ($q_{net,i}$) surfaces, are also shown in Fig. 11. Over the preheat region, the net heat flux at the inner surface ($q_{net,i}$) is more as compared to the net heat flux at outer surface due to the radiation heat gain at inner core region. It is observed that conduction heat flux at the inner and outer surfaces at the flame anchor location (X=0) is one order of magnitude larger than radiation heat fluxes; yet, gas radiation has an influence that extends further upstream. Thus, net heat flux at inner and outer surface ($q_{net,o}$ and $q_{net,i}$) follow closely conduction heat flux ($q_{c,o}$ and $q_{c,i}$) near flame anchor location (X=0) and further upstream it follows radiation heat flux at outer and inner surfaces ($q_{r,o}$ and $q_{r,i}$).

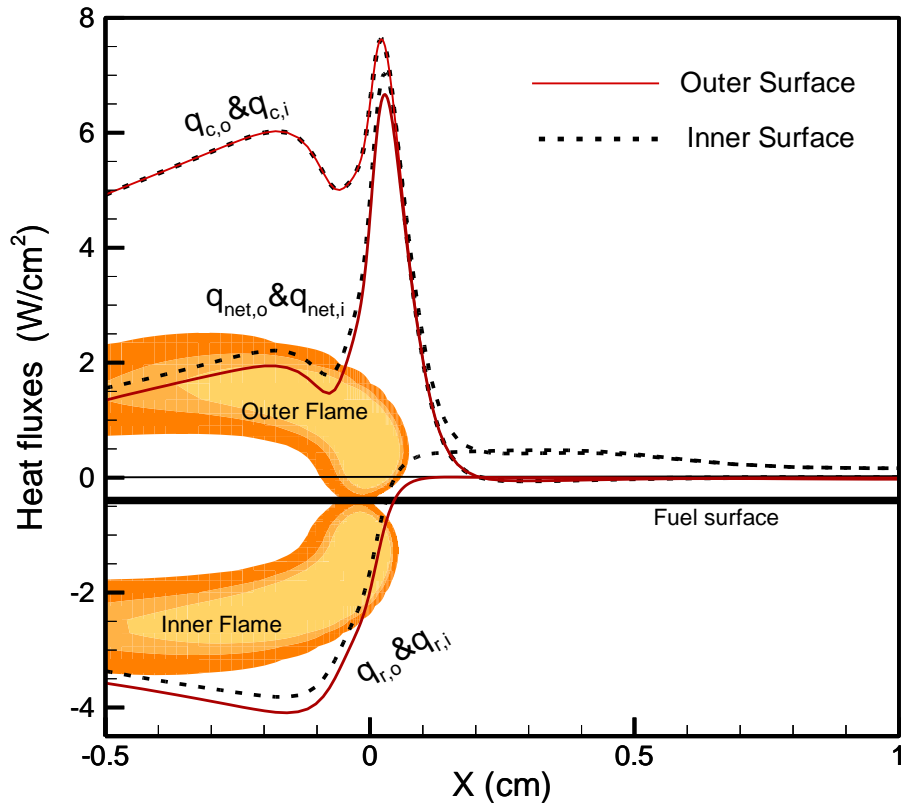

Fig. 11. Representation of different heat feedback at inner and outer surface of fuel along with flame (Temperature contour) at 14.5 mm fuel radius and 25 cm/s flow speed.

Heat flux to preheat region of solid surface controls the flame spread phenomena. Therefore, the different mode of heat fluxes over the preheat region for 14.5 mm, 19 mm and 24.5 mm fuel radius at outer surface (Fig. 12(a) and 12(b)) and inner surface (Fig. 12(c) and 12(d)) are compared in normal gravity environment. These heat fluxes are at 25 cm/s opposed flow speed. The insets in these figures show the comparison of heat fluxes of different fuel radius over a part of preheat region.

In the preheat region the net radiation and net heating value at the inner surface is higher as compared to net radiation and net heating value at the outer surface (Fig. 12 (a)- 12(d)), Also one can see in the insets of figures that the net radiation and net heating value at outer and inner surface for 24.5 mm fuel radius is higher as compared to 19 mm and 14.5 mm fuel radii. This indicates the increasing trend of flame spread rate with respect to fuel radii considered here.

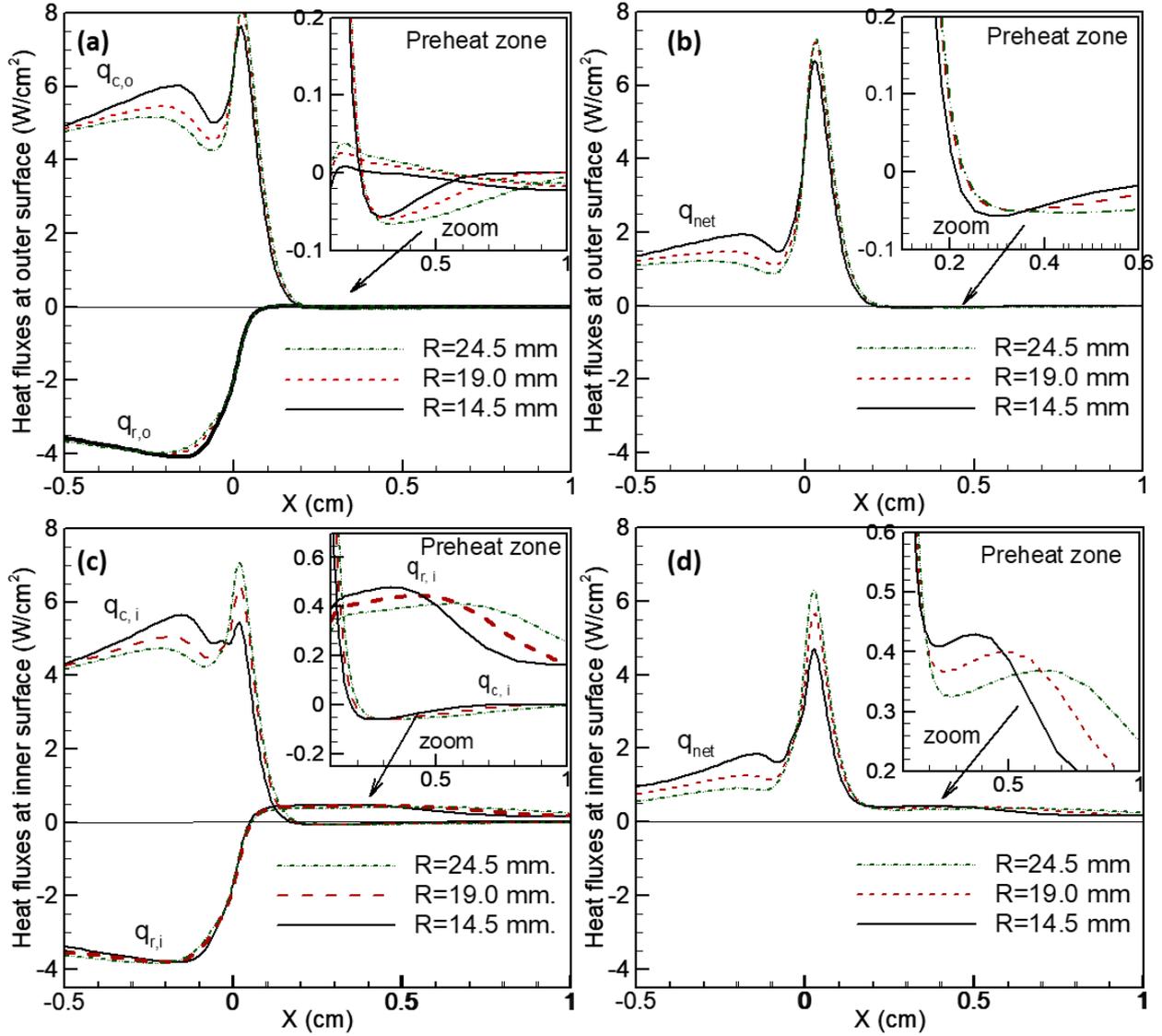

Fig.12. Heat fluxes over different fuel radius at 25 cm/s flow speed in normal gravity (a) conduction and radiation heats at outer surface (b) net heat flux at outer surface (c) conduction and radiation heats at inner surface (d) net heat flux at inner surface.

Furthermore, in normal gravity, a comparison of integrated heat fluxes is presented in Table 2 for different fuel radii. In Table 2, it is observed that, as the radius of the fuel increases from 14.5 mm to 24.5 mm, the integrated conduction heat flux per unit circumference ($\int_{x=0}^{\infty} q_c dx$) at outer surface increases from 9.59 W/m to 10.79 W/m and at inner surface, it increases from 5.78 W/m to 8.37 W/m. On other hand, with increase of radius from 14.5 mm to 24.5 mm, net radiation heat ($\int_{x=0}^{\infty} q_r dx$), loss at outer surface per unit circumference decreases from -3.43 W/m to -2.29 W/m (-ve signs are shown to indicate heat loss). however, net radiation heat gain at inner surface per unit circumference increases from 8.58 W/m to 10.16 W/m.

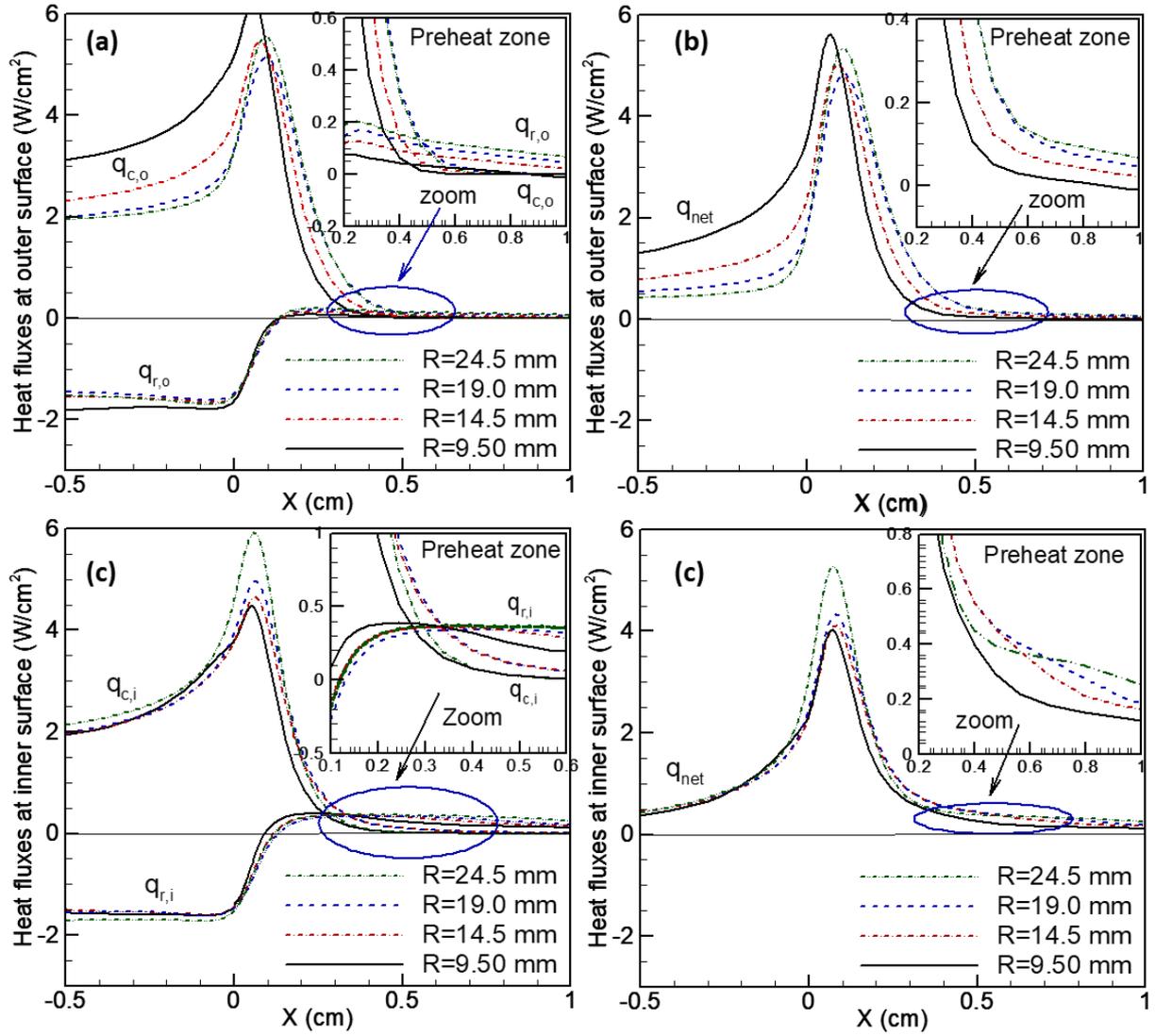

Fig.13. Heat fluxes over different fuel radius at 25 cm/s flow speed in micro gravity (a) conduction and radiation heats at outer surface (b) net heat flux at outer surface (c) conduction and radiation heats at inner surface (d) net heat flux at inner surface.

Note here the net radiation ($\int_{x=0}^{\infty} q_r dx$) at inner and outer surface is the summation of radiation heat loss from the solid surface ($\int_{x=0}^{\infty} q_{s,r} dx$) and radiation heat gain through gas phase ($\int_{x=0}^{\infty} q_{g,r} dx$) at inner and outer surfaces respectively. The different heat fluxes shown in Fig. 12 have only net radiation heat flux ($q_{ro}$ and $q_{ri}$) to make it distigushable between different fuel radii but the integrated values per unit circumference is calculated seperately for radiation heat loss from solid surface and radiation heat gain throgh gas phase in the Table 2.

The net heat flux $q_{net}$, is the sum of conduction and net radiation heat fluxes ($q_{net} = q_c + q_r$) at outer and inner surfaces and shown in Fig. 12 (b) and 12 (d) respectively. In the Fig. 12 (b), the net heat flux at outer surface is mainly due to the conduction heat flux and peaks near the flame anchor location (X=0) but radiation results in heat loss from preheat region of the outer fuel surface. On other hand net heat flux at inner surface (Fig. 12 (d)), over the preheat region show the heat gain due to radiation. The flame spread rate over the circular duct is directly proprtional to the net incident heating rate at preheat region. In normal gravity, as the radius of the fuel increases from 14.5 mm to 24.5 mm, the net heating rate increases from 20.57 W/m to 27.12 W/m in normal gravity, thus, increasing the flame spread rate from 7.9 mm/s to 10.4 mm/s (Table 2).

Similar to normal gravity, the different mode of heat fluxes over the preheat region for different fuel radii (9.5 mm, 14.5 mm, 19 mm and 24.5 mm) at outer surface (Fig. 13(a) and 13(b)) and inner surface (Fig. 13(c) and 13(d)) are compared in microgravity environment. These heat fluxes are plotted against flow speed of 25 cm/s. The insets in these figures show the enlarged heat fluxes of different fuel radii over a part of preheat region. The corresponding value of integrated heat flux values per unit circumference are also calculated over preheat region and presented in Table 3.

In Fig. 13, it can be observed that, in the preheat region near the flame anchor location the net heat flux for 24.5 mm fuel radius is highest followed by 19 mm, 14.5 mm and 9.50 mm fuel radii at inner and outer surface (insets in Fig. 13 (b) and 13 (d)). As the radius of the fuel increases from 9.5 mm to 24.5 mm, the conduction heat flux per unit circumference at outer surface ($\int_{x=0}^{\infty} q_{c,o} dx$) increases from 16.4 W/m to 18.46 W/m and at inner surface ($\int_{x=0}^{\infty} q_{c,i} dx$), it increases from 10.72 W/m to 13.54 W/m. On other hand with increase of radius from 9.5 mm to 24.5 mm, net radiation heat feedback loss at outer surface ($\int_{x=0}^{\infty} q_{r,o} dx$) per unit circumference decreases from -5.79 W/m to -2.23 W/m. however, radiation heat gain at inner surface ($\int_{x=0}^{\infty} q_{r,i} dx$) per unit circumference increases from 4.82 W/m to 6.79 W/m.

Similar to normal gravity, here also, the net radiation ($\int_{x=0}^{\infty} q_r dx$) at inner and outer surface is the summation of radiation heat loss from the solid surface ($\int_{x=0}^{\infty} q_{s,r} dx$) and radiation heat gain through gas phase ($\int_{x=0}^{\infty} q_{g,r} dx$) at inner and outer surfaces respectively. The different heat fluxes shown in Fig. 13 have only net radiation heat flux ($q_{ro}$ and $q_{ri}$) to make it distigushable with different fuel radii but the integrated values per unit circumference is calculated seperately for radiation heat loss from solid surface and radiation heat gain throgh gas in the Table 3.

The net heat flux $q_{net}$, is the sum of conduction and net radiation heat fluxes ($q_{net} = q_c + q_r$) at outer and inner surfaces and shown in Fig. 13 (b) and 13 (d) respectively. In the Fig. 13 (b), the net heat flux at outer surface is mainly due to the conduction heat flux and peaks near the flame anchor location (X=0) but net radiation heat flux results in heat loss from preheat region at outer fuel surface of the solid fuel (Table 3). On other hand net heat flux at inner surface (Fig. 13 (d)), over the preheat region (both from conduction and radiation heat fluxes) show the heat gain. As the radius of the fuel increases from 9.5 mm to 24.5 mm, in microgravity, the net heat flux curve at inner surface extending far upstream and the integrated value of radiation heat flux per unit circumference at inner surface increases from 15.56 W/m to 20.33 W/m.

| R mm | Surface | Component of integrated heat flux per unit circumference over preheat region in normal gravity (W/m) | | | | | | $V_f$ mm/s |
|---|---|---|---|---|---|---|---|---|
| | | $\int_{x=0}^{\infty} q_c dx$ | $\int_{x=0}^{\infty} q_{g,r} dx$ | $\int_{x=0}^{\infty} q_{s,r} dx$ | $\int_{x=0}^{\infty} q_r dx$ | $\int_{x=0}^{\infty} q_{net} dx$ | $\int_{x=0}^{\infty} (q_{net,i} + q_{net,o}) dx$ | |
| 14.5 | Inner | 5.78 | 14.69 | -6.1 | 8.58 | 14.37 | 20.57 | 7.9 |
| | Outer | 9.59 | 4.36 | -7.76 | -3.43 | 6.19 | | |
| 19.0 | Inner | 7.33 | 17.2 | -7.99 | 9.23 | 16.56 | 24.52 | 9.5 |
| | Outer | 10.65 | 5.12 | -7.85 | -2.73 | 7.96 | | |
| 24.5 | Inner | 8.37 | 19.46 | -9.29 | 10.16 | 18.59 | 27.12 | 10.4 |
| | Outer | 10.79 | 5.65 | -7.96 | -2.29 | 8.53 | | |

Table 2: Integrated heat flux values at inner and outer surfaces of different radius fuels in normal gravity at 25 cm/s flow speed

| R mm | Surface | Component of integrated heat flux per unit circumference over preheat region in microgravity (W/m) | | | | | | $V_f$ mm/s |
|---|---|---|---|---|---|---|---|---|
| | | $\int_{x=0}^{\infty} q_c dx$ | $\int_{x=0}^{\infty} q_{g,r} dx$ | $\int_{x=0}^{\infty} q_{s,r} dx$ | $\int_{x=0}^{\infty} q_r dx$ | $\int_{x=0}^{\infty} q_{net} dx$ | $\int_{x=0}^{\infty} (q_{net,i} + q_{net,o}) dx$ | |
| 9.50 | Inner | 10.72 | 10.72 | -5.87 | 4.82 | 15.56 | 26.22 | 11.2 |
| | Outer | 16.40 | 3.07 | -8.83 | -5.79 | 10.65 | | |
| 14.5 | Inner | 12.68 | 13.27 | -8.63 | 4.64 | 17.31 | 28.65 | 12.2 |
| | Outer | 17.04 | 4.18 | -9.87 | -5.69 | 11.34 | | |
| 19.0 | Inner | 12.65 | 15.63 | -10.34 | 5.29 | 17.94 | 32.31 | 13.8 |
| | Outer | 18.65 | 5.40 | -9.78 | -4.28 | 14.37 | | |
| 24.5 | Inner | 13.54 | 18.97 | -12.18 | 6.79 | 20.33 | 36.56 | 15.5 |
| | Outer | 18.46 | 7.26 | -9.49 | -2.23 | 16.23 | | |

Table 3: Integrated heat flux values at inner and outer surfaces of different radius fuels in microgravity at 25 cm/s flow speed

### 6.4. Flame spread rate over circular duct in normal gravity and microgravity- effect of fuel radius and opposed flow speed

In the experiments using the setup as shown in Fig. 3, the radius of the circular duct fuel varies from 5 mm to 24.5 mm and flow speed ranging from no flow (0 cm/s) to 30 cm/s in the normal gravity environment. The data obtained from the experiments and numerical calculations are shown in Fig. 14. The experimental data and numerical calculation both show that as the radius of circular duct increases the flame spread rate also increases at all flow speeds in normal gravity environment.

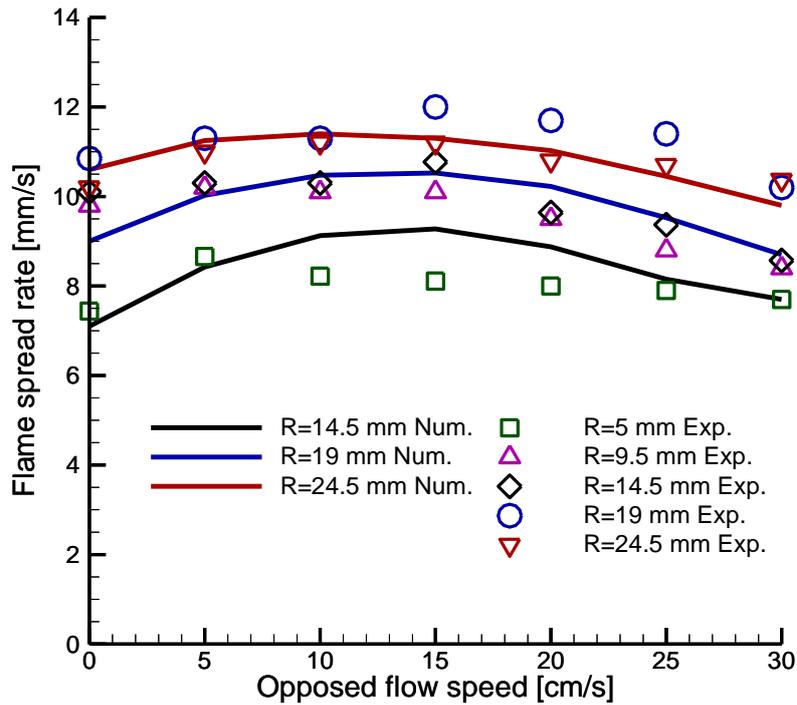

Fig.14. Flame spread rate with respect to opposed flow speed over different fuel radius in normal gravity obtained experimentally and numerically

One can note that numerical model presented in this study is able to predict the flame spread rate trend for 24.5 mm, 19 mm and 14.5 mm fuel radii however, the further lowering the fuel radius like 9.5 mm and 5 mm, the flame extinguishes in numerical model. In Fig. 10, as the opposed flow speed increases from no flow to 30 cm/s flow speed, initially the flame spread rate increases for all diameters and start decreasing. The initial increase of flame spread rate with increase in fuel radius is due to increase in oxygen supply at reaction zone of inner core region but further increase in flow speed in normal gravity cools the flame so flame spread rate starts decreasing. The non-monotonic increasing decreasing trend of flame spread rate with respect to flow speed follows the trend of maximum flame temperature and flame length which is discussed in earlier section 6.1 (Fig. 7).

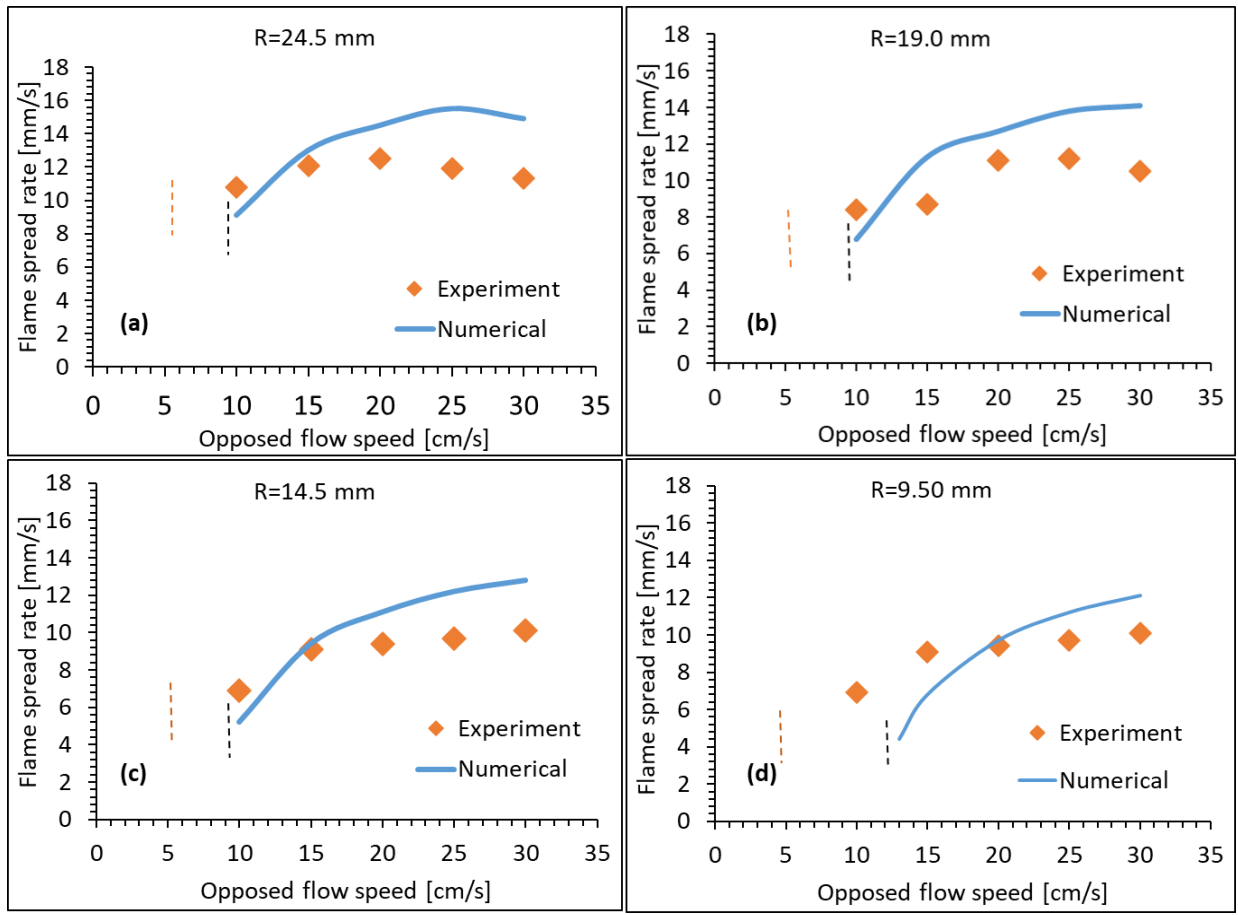

Fig. 15. Comparison of flame spread rate values obtained experimentally and computed from numerical model in microgravity (a) 24.5 mm fuel radius (b) 19.0 mm radius (c) 14.5 mm radius and (d) 9.50 mm radius.

In the drop experiments, the radius of the circular duct fuel varies from 5 mm to 24.5 mm and opposed flow speed varies from 5 cm/s to 30 cm/s in the microgravity environment. The data obtained from the experiments and numerical calculations for each fuel radius (9.5 mm, 14.5 mm, 19.0 mm and 24.5 mm) are shown in Fig. 15. In the present experiments as the flow speed decreases to 5 cm/s, the flame does not sustain in microgravity environment. A similar extinction is also obtained from the numerical calculations which is below 10 cm/s. In case of 9.5 mm fuel radius (Fig. 15 (d)) the extinction of flame in numerical model is at 13 cm/s flow speed however, in experiments a weak flame is obtained up to 10 cm/s flow speed. One can note here in both numerical model and experiments that as the flow speed increases, the flame spread rate shows non-monotonic increasing decreasing trend for fuel radius of 24.5 mm but monotonic increasing trend for fuel radii of 19.0 mm, 14.5 mm and 9.50 mm. Further increase in fuel radius on may observe the decrease in flame spread rate as the net radiation heat flux will decrease at larger fuel radii due to decrease the radiation heat feedback from the hot char.

Furthermore, similar to normal gravity, as the fuel radius decreases from 9.5 mm 24.5 mm, the flame spread rate increases. The increase of flame spread rate with increase in fuel radius is due to proportionally increase in heat feedback rate at inner and outer surface of the fuel (Table 3). Unlike normal gravity, in microgravity environment the pyrolysis and char length is higher. The hot burnt section of the fuel in downstream heat the unburnt section of circular duct fuel in inner core region by radiation. The radiation at inner core region increases with increase in fuel radius studied here and eventually enhances the flame spread rate.

### 6.5. Effect of uneven fuel mass flux distribution at inner and outer region

The present numerical simulation assumes even fuel-vapor mass flux distribution at inner and outer surfaces of the circular duct fuels. But it is interesting to see the effect of uneven mass flux distribution at inner and outer surfaces of the fuel. So, additional simulations considering the different mass flux of fuel vapor in normal gravity and microgravity is presented in Fig. 16. One can note that maximum flame spread rate in case of normal gravity environment (Fig. 16 (a)) is appears when 60% fuel vapor is coming though outer surface and 40 % reaching at inner core, irrespective of fuel radius. In case of microgravity environment, as the fuel radius increase the peak of flame spread rate towards left with increase in fuel radius (Fig. 16 (b)). The maximum flame spread for 9.5 mm fuel radius is appears when the 70% fuel vapor coming from outer surface (30% from inner surface) whereas, in the case of 24.5 mm fuel radius, the peak of flame spread appears when fuel vapor coming from outer surface at around 55%.

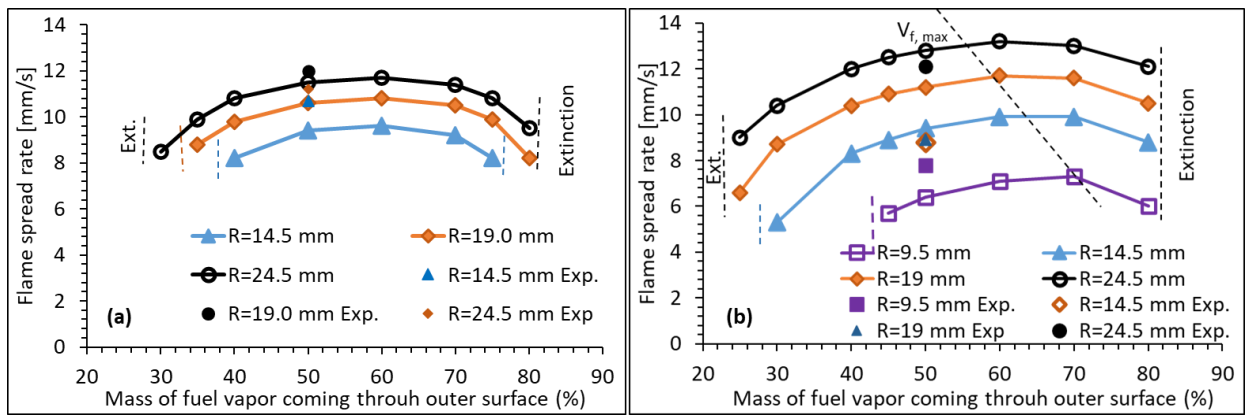

Fig. 16. Effect of uneven mass flux distribution at inner and outer fuel surface in (a) normal gravity and (b) microgravity environments for different fuel radius at 15 m/s flow speed

The discrepancy in the flame spread rate with respect to uneven mass flux distribution is due the pressure difference which appears higher at inner core region as compared to outer region of the fuel surface. The pressure difference in inner core region and outer region is due to the difference in velocity profile at inner and outer surface.

**6.6.    Effect of gas absorption coefficient on flame spread rate.**

In the previous section 6.3. It is observed that radiation heat flux is key parameter for flame spread rate over the thin circular duct fuels. There are two source of radiation flux over the preheat region, one is due to flame and other is due to hot chat in the downstream. In opposed flow flame spread thin fuels, the length of the flame is small therefor one can expect a small radiation contribution as compared to hot char. In order to bring the role of radiation flux from the hot char which is at temperature of 950 K in normal gravity and 800 K in microgravity, additional simulations have been carried out without absorption coefficient (Neglecting radiation flux from the flame). One can see in the fig. 17 (a) and (b) that, there is no significant change in the flame spread rate with and without considering the absorption coefficient. Here, the solid emissivity is taken as 0.92. The flame spread rate with respect to fuel diameter increases both in normal gravity and microgravity environments indicating the influence of char radiation at inner surface of the circular duct fuel.

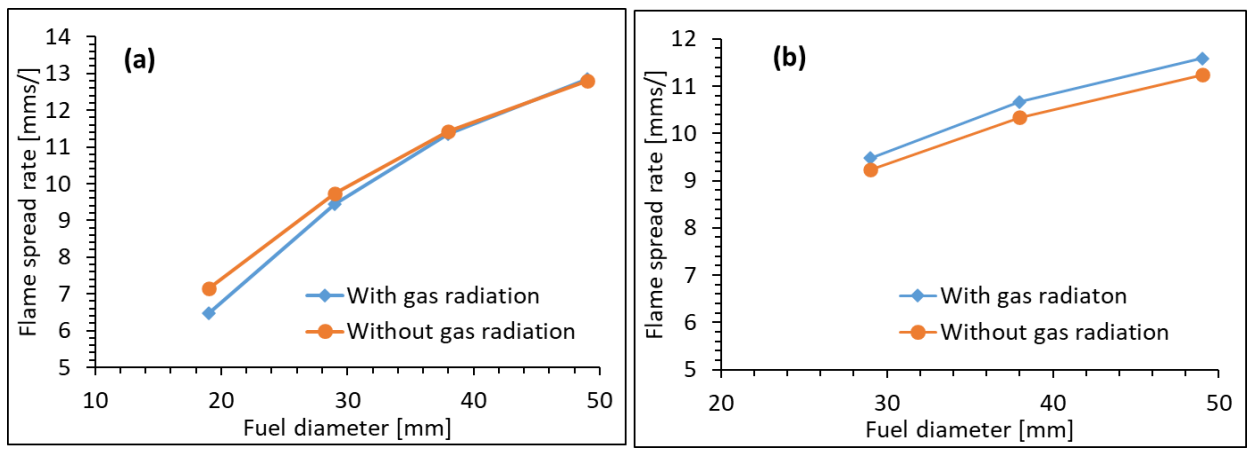

Fig. 17. Effect of flame radiation (a) in normal gravity and (b) microgravity for different fuel radius at 15 cm/s flow speed

**6.7. Effect of solid surface emissivity on flame spread rate.**

In previous section we have seen that radiation from hot solid surface has significant role in the flame spread rate. The effect of different emissivity of solid fuel is calculated in normal gravity and microgravity environment. Flame spread rate over three fuel radii, 14.5 mm, 19 mm and 24.5 mm is plotted at different value of emissivity in Fig. 18 (a). In normal gravity there is not any significant change of flame spread rate when solid emissivity changes between '0' to 1. In microgravity environment, the flame spread rate increases with increase in solid emissivity from '0' to 0.3 and the flame spread rate start decreasing when emissivity further increases from 0.3 to 1. This non-monotonic nature of flame spread rate with respect to solid emissivity

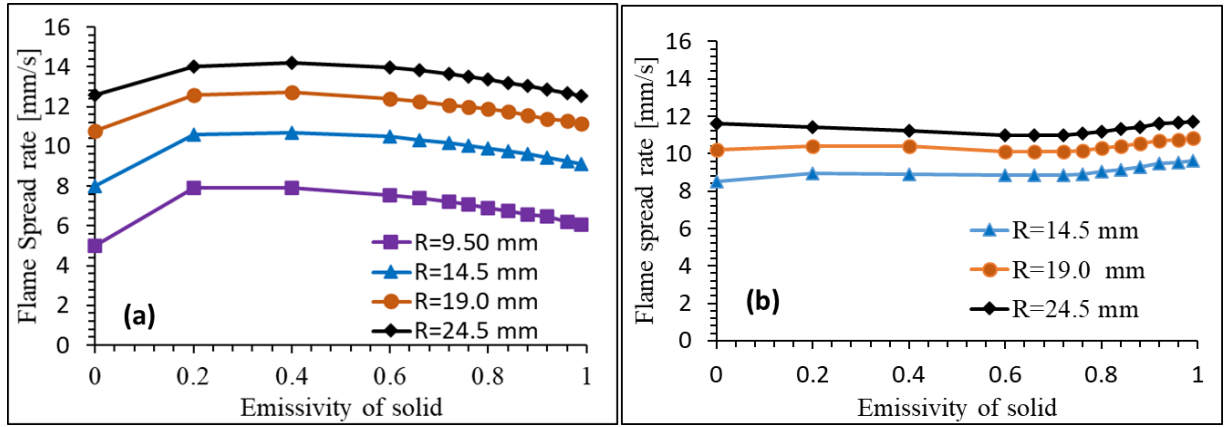

Fig. 18. Effect of solid surface emissivity on flame spread over different fuel radius of circular ducts in (a) normal gravity and (b) microgravity at 15 cm/s flow speed

## 7. Conclusion

The present work investigates the flame spread rate over circular duct in normal gravity and microgravity environments. Experiments as well as 2-D axisymmetric numerical model is used to understand the flame spread phenomena over circular duct at ambient oxygen concentration of 21 % and 1 atm. pressure. The study is performed over different fuel radii (5 mm, 9.5 mm, 14.5 mm, and 24.5 mm) and varying flow speeds ranging from no flow (0 cm/s) to 30 cm/s. The key observations of this study are as follows.

1. The flame length of circular duct geometry increases in normal gravity and microgravity with increase in fuel radius up to 24.5 mm fuel radius considered here.
2. For circular duct geometry, in normal gravity environment as the opposed flow speed increases the flame length shows non-monotonic increasing decreasing trend and there is no change of flame luminosity However, in case of microgravity the flame length and luminosity both increases with increase in opposed flow speed in present study.
3. The net radiation heat flux (summation of radiation loss from solid surface and radiation gain through gas phase) show loss of heat from the outer surface of circular duct and gain of radiant heat at inner surface. The gain of radiation heat at inner surface is due to the radiation heat feedback from the hot char.

4. As the radius of circular duct fuel increases the net heat flux over the circular duct increases which eventually increase the flame spread rate.